\documentclass[11pt]{article}

\usepackage{graphicx}
\usepackage{thumbpdf}
\usepackage{amsfonts,amstext,amsmath,amsthm}
\usepackage{accents}
\usepackage{color}
\usepackage{rotating}
\usepackage[utf8]{inputenc}
\usepackage{booktabs}
\usepackage{tikz}
\usepackage{multirow}
\usepackage{float}
\usepackage{natbib}
\usepackage{hyperref}
\usepackage[margin=2.5cm]{geometry}
\usepackage{array}
\usepackage[labelfont=bf, width=15cm, font=small]{caption}
\newcommand*{\doi}[1]{\href{http://dx.doi.org/#1}{doi: #1}}

\usepackage{accents}

\newcommand{\norm}[1]{\left\lVert#1\right\rVert}
\newcommand{\nnorm}[1]{\lVert#1\rVert}

\newcommand{\rZ}{Z}
\newcommand{\rY}{Y}
\newcommand{\rX}{\mX}

\newcommand{\rz}{z}
\newcommand{\ry}{y}
\newcommand{\rx}{\xvec}

\newcommand{\pZ}{F_\rZ}

\newcommand{\pN}{\Phi}
\newcommand{\pSL}{F_{\SL}}
\newcommand{\pMEV}{F_{\MEV}}

\newcommand{\dZ}{f_\rZ}
\newcommand{\dY}{f_\rY}

\newcommand{\h}{h}

\newcommand{\bernx}[1]{\bvec_{\text{Bs},#1}}

\newcommand{\parm}{\varthetavec}
\newcommand{\eparm}{\vartheta}

\newcommand{\shiftparm}{\betavec}

\newcommand{\ie}{\textit{i.e.,}~}
\newcommand{\eg}{\textit{e.g.,}~}
\newcommand{\cf}{\textit{cf.}~}

\newcommand{\Prob}{\mathbb{P}}
\newcommand{\Ex}{\mathbb{E}}
\newcommand{\RR}{\mathbb{R}}

\usepackage{dsfont}
\newcommand{\I}{\mathds{1}}

\newcommand{\given}{\vert}

 \DeclareMathOperator{\expit}{expit}

 \DeclareMathOperator*{\argmin}{{arg\,min}}

 \DeclareMathOperator{\SL}{SL}
 \DeclareMathOperator{\MEV}{MEV}

\def \avec {\text{\boldmath$a$}}    \def \mA {\text{$\mathbf A$}}
\def \bvec {\text{\boldmath$b$}}    \def \mB {\text{$\mathbf B$}}

\def \evec {\text{\boldmath$e$}}

    \def \mM {\text{$\mathbf M$}}

\def \rvec {\text{\boldmath$r$}}

\def \xvec {\text{\boldmath$x$}}    \def \mX {\text{$\mathbf X$}}
\def \yvec {\text{\boldmath$y$}}

\def \rA {\text{\boldmath$A$}}

\def \rH {\text{\boldmath$H$}}

\def \rM {\text{\boldmath$M$}}

\def \rX {\text{\boldmath$X$}}

 \def \calF {\mathcal F}

 \def \calN {\mathcal N}

\def \betavec         {\text{\boldmath$\beta$}}

\def \deltavec        {\text{\boldmath$\delta$}}

\def \varepsilonvec   {\text{\boldmath$\varepsilon$}}
\def \zetavec         {\text{\boldmath$\zeta$}}

\def \thetavec        {\text{\boldmath$\theta$}}
\def \varthetavec     {\text{\boldmath$\vartheta$}}

\def \muvec           {\text{\boldmath$\mu$}}
\def \nuvec           {\text{\boldmath$\nu$}}

\newcommand{\ubar}[1]{\underaccent{\bar}{#1}}

\newcommand{\pkg}[1]{\textbf{#1}}
\newcommand{\prm}{\mathbf{\Pi}_{\mA}}

\newcommand{\bv}{\betavec}

\newcommand{\ve}{\varepsilon}
\newcommand{\veX}{\ve_\rX}
\newcommand{\veY}{\ve_\rY}

\newcommand{\veA}{\ve_\rA}

\newcommand{\proglang}[1]{\textsf{#1}}
\newcommand{\code}[1]{\texttt{#1}}

\DeclareMathOperator{\Id}{Id}

\DeclareMathOperator{\pdo}{do}
\DeclareMathOperator{\push}{push}
\DeclareMathOperator{\Rade}{Rademacher}

\DeclareMathOperator{\Corr}{Corr}

\usepackage{tikz}

\usetikzlibrary{positioning, calc, shapes.geometric, shapes.multipart,
  shapes, arrows.meta, arrows,
  decorations.markings, external, trees, automata}
\tikzstyle{line} = [draw, -latex']
\tikzstyle{Arrow} = [
        thick,
        decoration={
                markings,
                mark=at position 1 with {
                        \arrow[thick]{latex}
} },
        shorten >= 3pt, preaction = {decorate}
        ]

\setcitestyle{authoryear,open={(},close={)}}

\newtheorem{definition}{Definition}
\newtheorem{example}{Example}
\newtheorem{proposition}{Proposition}

\usepackage{color}

\begin{document}

\title{\bf Distributional Anchor Regression}

\author{Lucas Kook$^{1,2}$%
        \footnote{Corresponding author, email: lucasheinrich.kook@uzh.ch \newline 
        Preprint; under review. Version: 08.04.2022. Licensed under CC-BY.} ,
        Beate Sick$^{1,2}$,
        Peter B{\"{u}}hlmann$^3$
}

\date{\small
$^1$University of Zurich, Switzerland $^2$Zurich University of Applied Sciences, Switzerland \\
$^3$ETH Zurich, Switzerland}

\maketitle

\begin{abstract}
Prediction models often fail if train and test data do not stem from the same distribution.
Out-of-distribution (OOD) generalization to unseen, perturbed test data is 
a desirable but difficult-to-achieve property for prediction models and in general requires 
strong assumptions on the data generating process (DGP).
In a causally inspired perspective on OOD generalization, the test data arise 
from a specific class of interventions on exogenous random variables of the DGP, 
called anchors.
Anchor regression models, introduced by \citet{rothenhaeusler2018anchor},
protect against distributional shifts in the test data by employing causal regularization.
However, so far anchor regression has only been used with a squared-error loss which is inapplicable 
to common responses such as censored continuous or ordinal data.
Here, we propose a distributional version of anchor regression which generalizes
the method to potentially censored responses with at least an ordered sample space.
To this end, we combine a flexible class of parametric transformation models 
for distributional regression with an
appropriate causal regularizer under a more general notion of residuals.
In an exemplary application and several simulation scenarios we demonstrate the
extent to which OOD generalization is possible.
\end{abstract}

\section{Introduction} \label{sec:intro}

Common methods in supervised statistical learning assume the test data to follow
the same distribution as the training data. 
This is implicitly exploited in, \eg cross-validation or by randomly 
splitting a dataset into a training and a test set, which has been demonstrated
to be potentially flawed \citep{efron2020prediction} 
due to concept drift or domain shift where new (test) data do not follow
the same distribution as the training data.
More recently, the problem has been referred to as out-of-distribution (OOD)
generalization \citep{arjovsky2019invariant}.
The desire to achieve reliable test predictions under distributional shifts is
ubiquitous in many fields of machine learning and statistics, such as
transfer learning \citep{pan2009survey,rojas2018invariant},
domain adaptation \citep{magliacane2018domain}, 
multi-task learning \citep{caruana1997multitask},
representation learning \citep{mitrovic2020representation} or 
prediction models in medical statistics \citep{subbaswamy2020development}.
Accordingly, many different formulations of the problem of OOD generalization
exist in the literature \citep[a detailed overview can be found in][]{chen2020domain}.
We will frame OOD generalization as the problem of robustly predicting an
outcome in novel, unseen environments, based on data from a few observed
environments and extend on the idea of anchor regression and causal regularization
\citep{rothenhaeusler2018anchor,buehlmann2018invariance,buhlmann2020deconfounding}
to develop distributional anchor regression.
In such a framework, training a model on heterogeneous data is not a disadvantage
but rather a necessity.

\subsection{Related work} \label{sec:related}

It has been known for decades that a causal model is robust towards
arbitrarily strong perturbations on components other than the response 
\citep{haavelmo1943statistical}. 
However, identifying causal structures is not only difficult but employing them for
prediction often leads to sub-par prediction performance when the test data 
contain perturbations of bounded strength \citep{rothenhaeusler2018anchor}.
\citeauthor{rothenhaeusler2018anchor} introduce linear anchor regression, 
which allows a trade-off between prediction performance and robustness
against shift perturbations of a certain size. 
The framework of linear anchor regression was extended to deal with nonlinear 
regression between the response and covariates \citep{buehlmann2018invariance}.
Furthermore, \citet{christiansen2020causal} provide a causal framework to decide 
which assumptions are needed for and to what extent OOD generalization 
is possible.

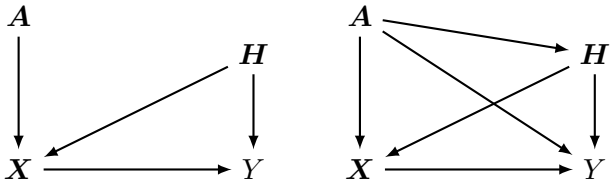
\begin{figure}[!ht]
\centering
\begin{tikzpicture}
 \node (X) {$\rX$};
 \node [above=1.5cm of X] (E) {$\rA$};
 \node [right=2.5cm of X] (Y) {$\rY$};
 \node [above =of Y] (H) {$\rH$};
 \draw[Arrow] (E) -- (X);
 \draw[Arrow] (X) -- (Y);
 \draw[Arrow] (H) -- (X);
 \draw[Arrow] (H) -- (Y);
\end{tikzpicture}
\hspace{0.5cm}
\begin{tikzpicture}
 \node (X) {$\rX$};
 \node [above=1.5cm of X] (E) {$\rA$};
 \node [right=2.5cm of X] (Y) {$\rY$};
 \node [above =of Y] (H) {$\rH$};
 \draw[Arrow] (E) -- (X);
 \draw[Arrow] (X) -- (Y);
 \draw[Arrow] (H) -- (X);
 \draw[Arrow] (H) -- (Y);
 \draw[Arrow] (E) -- (H);
 \draw[Arrow] (E) -- (Y);
\end{tikzpicture}
\caption{Graphical models for the response variable $\rY$, covariates $\rX$ and hidden 
confounders $\rH$: IV regression with instruments $\rA$ (left) 
and anchor regression with anchor $\rA$ (right).
In anchor regression, $\rA$ is only required to be a source node but is allowed to directly influence
response, covariates and hidden confounders.
} \label{fig:dag3}
\end{figure}

Anchor regression is related to Instrumental Variables (IV) regression.
However, the main IV assumption that the instrument $\rA$ does not directly 
affect some hidden confounding variables $\rH$ is dropped, at the price of 
non-identifiability of the causal parameter \citep{angrist1996identification}.
A graphical description of the issue is given in Figure~\ref{fig:dag3}.

\subsection{Our Contribution} \label{sec:contrib}

In this work we develop a framework for  
distributional anchor regression in the broad class of  
transformation models \citep[TMs,][]{hothorn2014conditional}.
The resulting class of anchor TMs generalizes (non-) linear anchor regression 
to potentially censored responses and characterizes the full conditional distribution 
of $\rY \given \rX = \rx$ instead of estimating solely the conditional mean function.
While the linear $L_2$ anchor loss can be decomposed into a squared error
and causal regularization term penalizing correlation between anchors and residuals, 
we propose a distributional anchor loss 
based on the negative log-likelihood and replacing the least-squares residuals by 
the more general score residuals.
The proposed causal regularizer induces uncorrelatedness between the anchors 
and these score residuals.
The resulting procedure is tailored towards protecting against distributional 
shifts induced by the anchors 
and naturally interpolates between the unpenalized maximum-likelihood
estimate and a solution for which anchors and residuals are strictly uncorrelated.
The latter may be thought of as a distributional IV-like objective 
but it generally does not estimate the causal model due to the fact that the anchor
$\rA$ can also directly influence $\rH$ and $\rY$ (see Figure~\ref{fig:dag3}) 
and the conditional expectation function $\Ex[\rY\given\rX]$ is generally non-linear.
It leads to some invariance of the score residuals across the values
of the anchors $\rA$, and such an invariance property has been referred to
as ``diluted causality'' \citep{buehlmann2018invariance}.

The rest of this paper is organized as follows.
In Section~\ref{sec:bg}, we present the necessary theoretical background for linear
$L_2$ anchor regression, transformation models and score residuals. Section~\ref{sec:danchor}
introduces distributional anchor regression and theoretical results on residual invariance
and identifiability of the causal parameter in instrumental variable settings.
Empirical results are presented in Section~\ref{sec:empirical}. We end with a discussion
of our contribution in Section~\ref{sec:discussion}.
In the appendix we present further details on notation, computation, score residuals
and further empirical results on point-prediction performance of distributional 
anchor regression in comparison with linear $L_2$ and non-linear anchor boosting.
We implement all methods and algorithms
in the \proglang{R} language for statistical computing 
\citep{pkg:base} and the code is available on 
\href{https://github.com/LucasKookUZH/distributional-anchor-regression.git}{GitHub}.

\section{Background} \label{sec:bg}

First, we introduce structural equation models (SEMs) before 
recapping linear anchor regression.
In Section~\ref{sec:tram}, we switch perspectives from modelling the conditional 
expectation to trans\-formation models which enable
to capture entire conditional distributions.
The notation used in this work is described in Appendix~\ref{app:notation}.

\subsection{Structural Equation Models} \label{sec:framework}

Let $\rY$ be a response which takes values in $\RR$, $\rX$ be a random vector of covariates 
taking values in $\RR^p$, $\rH$ denotes hidden confounders with sample space $\RR^d$, 
and $\rA$ are exogenous variables (called anchors, due to exogeneity; 
source node in the graph in Figure \ref{fig:dag3}) taking values in $\RR^q$.
The SEM governing linear anchor regression is given by
\begin{align} \label{eq:sem}
    \begin{pmatrix}
    \rY \\ \rX \\ \rH
    \end{pmatrix}
    \leftarrow
    \mB
    \begin{pmatrix}
    \rY \\ \rX \\ \rH
    \end{pmatrix}
    + \mM \rA + \varepsilonvec,
\end{align}
with $(1+p+d) \times (1+p+d)$-matrix $\mB$ 
which corresponds to the structure of the SEM in terms of a directed acyclic graph (DAG),
the effect of $\rA$ enters linearly via the $(1 + p +d ) \times q$-matrix $\mM$, and 
$\varepsilonvec$ denotes the error term with mutually independent components. 
The ``$\leftarrow$'' symbol is algebraically a distributional equality sign.
It emphasizes the structural character of the SEM, saying that, \eg $\rY$ is only 
a function of the parents of the node $\rY$ in the structural DAG and the first
entry in the additive component $(\rM \rA + \varepsilonvec)$.

The anchors $\rA$ may be continuous or discrete. 
In the special case of discrete anchors each level can be viewed as an ``environment''.

We define perturbations as intervening on $\rA$, \eg
by $\pdo(\rA = \avec)$,
which replaces $\rA$ by $\avec$ in the SEM while leaving the underlying
mechanism, \ie the coefficients in the SEM, unchanged.
In this work we restrict ourselves to $\pdo$- \citep{pearl2009causality} and 
$\push$-interventions \citep{markowetz2005probabilistic}
on $\rA$, which in turn lead to shifts in the distribution of $\rX$.
In essence, $\pdo$-interventions replace a node in a graph with a deterministic
value, whereas $\push$-interventions are stochastic and only ``push'' the distribution
of the intervened random variable towards, \eg having a different mean.
Since $\rA$ is exogenous and a source node in the graph, the specific type of
intervention does not play a major role.
\citet{christiansen2020causal} show that under the above conditions
OOD generalization is possible in linear models.

\subsection{Linear Anchor Regression} \label{sec:lin-anchor}

Linear $L_2$ anchor regression with its corresponding causal regularization estimates the 
linear regression parameter $\bv$ as
\begin{align*}
\hat{\bv} = \argmin_{\bv} \bigg\{ \norm{(\Id - \prm)(\yvec -\mX\bv)}_2^2/n +
    \gamma\norm{\prm(\yvec - \mX\bv)}_2^2/n \bigg\},
\end{align*}
where $0 \le \gamma \le \infty$ is a regularization parameter and 
$\prm = \mA(\mA^\top\mA)^{-1}\mA^\top$ denotes the orthogonal projection onto
the column space of the anchors \citep{rothenhaeusler2018anchor}.
For $\gamma = 1$ one obtains ordinary least squares, $\gamma \to \infty$ corresponds
to two-stage least squares as in instrumental variables regression and $\gamma = 0$
is partialling out the anchor variables $\rA$ (which is equivalent to ordinary least 
squares when regressing $\rY$ on $\rX$ and $\rA$).
Causal regularization encourages, for large values of $\gamma$, uncorrelatedness of 
the anchors $\rA$ and the residuals.
As a procedure, causal regularization does not depend at all on
the SEM in eq.~\eqref{eq:sem}.
However, as described below, the method inherits a distributional robustness property,
whose formulation depends on the SEM in eq.~\eqref{eq:sem}.%

\citet{rothenhaeusler2018anchor} establish the duality between the $L_2$ loss in 
linear anchor regression and optimizing a worst case risk over specific shift perturbations.
The authors consider shift perturbations $\nuvec$, which are confined to be in
the set
\begin{align*}
C_\gamma := \big\{\nuvec : \nuvec = \mM\deltavec, \; 
\deltavec \mbox{ independent of } \varepsilonvec, \;
\Ex[\deltavec\deltavec^\top] \preceq \gamma \Ex[\rA\rA^\top]\big\},
\end{align*}
and which generate the perturbed response $\rY^\nuvec$, and covariates $\rX^\nuvec$ via
\begin{align*}
    \begin{pmatrix}
    \rY^\nuvec \\ \rX^\nuvec \\ \rH^\nuvec
    \end{pmatrix}
    \leftarrow
    \mB
    \begin{pmatrix}
    \rY^\nuvec \\ \rX^\nuvec \\ \rH^\nuvec
    \end{pmatrix}
    + \nuvec + \varepsilonvec.
\end{align*}
The set $C_\gamma$ contains all vectors which lie in the span of the columns 
of $\mM$ and thus in the same direction as the exogenous contribution $\mM \rA$ 
of the anchor variables.
The average size and direction of the perturbations $\deltavec$ is limited by $\gamma$
and the centered anchors' variance-covariance matrix.
Now, the explicit duality between the worst case risk over all shift perturbations
of limited size and the linear $L_2$ anchor loss is given by
\begin{align} \label{eq:anchorpop}
  \sup_{\nuvec \in C_\gamma} \Ex\left[(\rY^\nuvec - (\rX^\nuvec)^\top\shiftparm)^2\right] 
  = 
  \Ex\left[((\Id - P_\rA)(\rY -\rX^\top\bv))^2\right] +
  \gamma\Ex\left[(P_\rA(\rY - \rX^\top\bv))^2\right],
\end{align}
where $P_\rA = \Ex[\cdot \given \rA]$ is the population analogue of $\prm$.
We note that the right-hand side is the population analogue of the objective 
function in anchor regression. Hence, causal regularization in anchor regression 
provides guarantees for optimizing worst-case risk across a class of shift perturbations.
The details are provided in \citet{rothenhaeusler2018anchor}.

\subsection{Transformation Models} \label{sec:tram}

We now switch perspective from models for the conditional mean to modelling conditional distributions.
Specifically, we consider transformation models \citep{hothorn2014conditional}.
TMs decompose the conditional distribution of $\rY \given \rx$ 
into a pre-defined distribution function $\pZ$, with log-concave density $\dZ$, 
and a (semi-) parametric transformation function $\h(\ry \given \rx)$,
which is monotone non-decreasing in $\ry$
\begin{align*}
    F_{\rY \given \rx}(\ry \given \rx) = \pZ(\h(\ry \given \rx)).
\end{align*}
This way, the problem of estimating a conditional distribution simplifies to
estimating (the parameters of) the transformation function $\h = \pZ^{-1} \circ F_{\rY \given \rx}$ 
(since $\pZ$, called inverse link, is pre-specified and parameter-free).
Depending on the complexity of $\h$, very flexible conditional distributions can be modelled.
\citet{hothorn2018most} give theoretical guarantees for the existence and uniqueness
of the transformation function $\h$ for absolute continuous, count and ordered discrete 
random variables. 
For the sake of generality, $\h$ is parametrized in terms of a basis expansion 
in the argument $\ry$ which can be as simple as a linear function in $\ry$ or as 
complex as a basis of splines to approximate a smooth function in $\ry$.

In this work, we assume the transformation function for a continuous response
can be additively decomposed into a linear predictor in $\rx$ and a smooth 
function in $\ry$ which is modelled as a Bernstein polynomial of order $P$ with 
parameters $\thetavec \in \RR^{P+1}$ \citep{hothorn2018most}, such that
$\h(y \given \rx) = \bernx{P}(\ry)^\top\thetavec + \rx^\top \beta$. 
Monotonicity of $\bernx{P}(\ry)^\top\thetavec$ and thereby of 
$\h(\ry\given\rx)$ can then be enforced via the $P$ linear constraints
$\theta_1 \leq \theta_2 \leq \dots \theta_{P+1}$. 
In case of an ordinal response taking values in $\{\ry_1, \ry_2, \dots, \ry_K\}$,
the transformation function is a monotone increasing step function, 
$\h(\ry_k \given \rx) = \theta_k + \rx^\top\shiftparm$, for $k = 1, \dots, K - 1$ and 
the additional constraint $\theta_K = + \infty$.
We summarize a transformation model based on its inverse link function $\pZ$,
basis $\bvec$, which may include covariates, and parameters $\parm$,
such that $F_{\rY \given \rx}(\ry \given \rx) = \pZ(\bvec(\ry,\rx)^\top\parm)$.
For instance, for a transformation model with continuous response and explanatory 
variables $\rx$ we thus use $\bvec(\ry,\rx) = (\bernx{P}(\ry)^\top, \rx^\top)^\top$ and
$\parm = (\thetavec^\top, \shiftparm^\top)^\top$, 
yielding $\h(\ry \given \rx) = \bernx{P}(\ry)^\top\thetavec + \rx^\top\shiftparm$.
For a TM with ordinal response we substitute the Bernstein basis with
a dummy encoding of the response, which we denote by $\tilde\yvec$ 
\citep[\eg][]{kook2020ordinal}.
Also note that the unconditional case is covered by the above formulation as well,
by omitting all explanatory variables from the TM's basis.

\begin{figure}[!ht]
\centering
\includegraphics[width=0.6\textwidth]{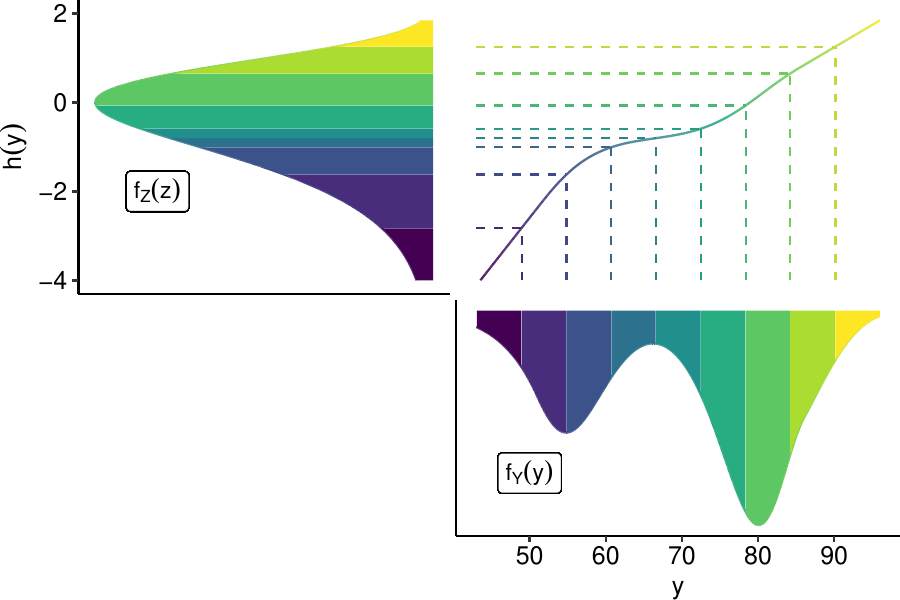}
\caption{
Illustration of an unconditional transformation model 
$(1 - \exp(-\exp(\cdot)), \bernx{6}, \parm)$ for 
the Old Faithful Geyser data \citep{azzalini1990look} using a 
Bernstein polynomial basis expansion of order 6 for the transformation function,
$\h(\ry) = \bernx{6}(\ry)$.
The colored regions indicate the transport of probability mass from 
$\Prob_Y$ (lower right) to $\Prob_Z$ (upper left) via the transformation function 
$\h(\ry) = \bvec(\ry)^\top\parm$ (upper right).
If $\h$ is continuously differentiable, the density of $\rY$ is given by
$\dY(\ry) = \dZ(\h(\ry))\h'(\ry)$.
}\label{fig:tram}
\end{figure}
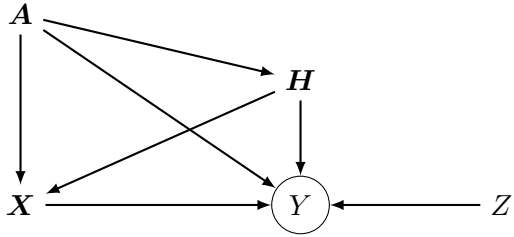
\begin{figure}[!ht]
\centering
\begin{tikzpicture}
 \node (X) {$\rX$};
 \node [above=2cm of X] (E) {$\rA$};
 \node[circle,draw] [right=3cm of X] (Y) {$\rY$};
 \node [above =of Y] (H) {$\rH$};
 \node [right =2cm of Y] (Z) {$\rZ$};
 \draw[Arrow] (E) -- (X);
 \draw[Arrow] (X) -- (Y);
 \draw[Arrow] (H) -- (X);
 \draw[Arrow] (H) -- (Y);
 \draw[Arrow] (E) -- (H);
 \draw[Arrow] (E) -- (Y);   
 \draw[Arrow] (Z) -- (Y);   
\end{tikzpicture}
\caption{
Structural equation model for a transformation model.
Instead of setting up the SEM on the scale of $\rY$, it is defined
on the scale of the inverse transformation function $\h^{-1}$.
The conditional distribution of $\rY := \h^{-1}(Z \given \rX, \rH, \rA)$ is still fully 
determined by $\h$ and $\pZ$.
The circle around $\rY$ emphasizes that its distribution is a deterministic function 
of its parents.
}
\label{fig:tsem}
\end{figure}

Figure~\ref{fig:tram} illustrates the intuition behind transformation models.
The transformation function (upper right panel) transforms the complex, bimodal
distribution of $\rY$ (lower panel) to $\pZ = \pMEV$, the standard minimum extreme
value distribution (upper left panel). 
An analogous figure for ordinal outcomes is published in \citet[Fig.~1]{kook2020ordinal}.
\begin{definition}[Transformation model, Definition~4 in \citet{hothorn2018most}]
The triple \\ ($\pZ$, \bvec, \parm) is called transformation model.
\end{definition}
\begin{example}[Linear regression]
The normal linear regression model (Lm) is commonly formulated as 
$\rY = \beta_0 + \rx^\top\tilde\shiftparm + \varepsilon$, $\varepsilon \sim \mathcal{N}(0, \sigma^2)$,
or $\rY \given \rx \sim \calN(\beta_0 + \rx^\top\tilde\shiftparm, \sigma^2).$
For a distributional treatment we write the above expression as
\begin{align} \label{eq:lm}
  F_{\rY \given \rx}(\ry \given \rx) 
  = \Phi\left(\frac{\ry - \beta_0 - \rx^\top\tilde\shiftparm}{\sigma}\right)
  = \Phi(\eparm_1 + \eparm_2 \ry - \rx^\top\shiftparm),
\end{align}
which can be understood as a transformation model 
by letting $\eparm_1 = - \beta_0 / \sigma$, $\eparm_2 = 1/\sigma$ and
$\shiftparm = \tilde\shiftparm / \sigma$. Formally, it corresponds to the model
\begin{align*}
(\pZ, \bvec, \parm) = \left(\Phi, (1, \ry, \rx^\top)^\top, (\eparm_1, \eparm_2, -\shiftparm^\top)^\top\right).
\end{align*}
Note that the baseline transformation, $\h(\ry \given \rX = 0)$, is constrained
to be linear with constant slope $\eparm_2$. 
Due to the linearity of $\h$ and the choice $\pZ=\Phi$, the modeled distribution of 
$\rY\given\rx$ will always be normal with constant variance.
By parametrizing $\h$ in a smooth way, we arrive at much more flexible conditional
distributions for $\rY \given \rx$.
\end{example}
The parameters of a TM can be jointly estimated using maximum-likelihood.
The likelihood can be written in terms of the inverse link function $\pZ$, 
which makes its evaluation computationally more convenient.
For a single datum $(\ry, \rx)$ with potentially censored response $\ry \in (\ubar\ry, \bar\ry]$
the log-likelihood contribution is given by \citep{lindsey1996parametric}
\begin{align*}
\ell(\parm; \ry, \rx) = \begin{cases}
    \log \dZ(\bvec(\ry,\rx)^\top\parm) + \log\left(\bvec'(\ry,\rx)^\top\parm \right), & 
        \text{exact,} \\
    \log \pZ(\bvec(\bar{\ry},\rx)^\top\parm), & 
        \text{left,}\\
    \log \left( 1 - \pZ(\bvec(\ubar{\ry},\rx)^\top\parm) \right), & 
        \text{right,} \\
    \log \left( \pZ(\bvec(\bar{\ry},\rx)^\top\parm) - \pZ(\bvec(\ubar{\ry},\rx)^\top\parm) \right), & 
        \text{interval.}
\end{cases}
\end{align*}
The likelihood is always understood as conditional on $\rX$ when viewing the covariables as random.
Allowing for censored observations is of practical importance, because
in many applications the response of interest is not continuous or suffers from
inaccuracies, which can be taken into account via uninformative censoring.
\begin{example}[Lm, cont'd]
For an exact datum $(y, \rx)$ the log-likelihood in the normal
linear regression model is given by 
\begin{align*}
  \ell(\eparm_1, \eparm_2, \shiftparm; \ry, \rx) 
     = \log \phi\big(\eparm_1 + \eparm_2 \ry - \rx^\top \shiftparm\big)
     + \log(\eparm_2),
\end{align*}
using the density approximation to the likelihood \citep{lindsey1996parametric}.
Here, $\phi$ denotes the standard normal density, 
and $\bvec'(\ry,\rx)^\top\parm = \frac{\partial \bvec(\ry,\rx)^\top\parm}{\partial \ry} = \eparm_2$.
\end{example}
Now that we have established TMs and the log-likeli\-hood function to
estimate their parameters, we also need a more general notion of the residuals to
formulate a causal regularizer for a distributional anchor loss. 
Most importantly, these residuals have to fulfill the same
requirements as least squares residuals in the linear $L_2$ anchor loss.
That is, they have to have zero expectation and a positive definite covariance
matrix \citep[Theorem~3 in][]{rothenhaeusler2018anchor}.
In the survival analysis literature, score residuals 
have received considerable attention, and fulfill the above requirements at least
asymptotically
\citep{lagakos1981residuals,barlow1988residuals,therneau1990martingale,farrington2000residuals}.
We now define score residuals for the general class of transformation models.
\begin{definition}[Score residuals] \label{def:score}
Let $(\pZ, \bvec, \hat\parm)$ be a TM with maximum-likelihood estimate $\hat\parm$.
On the scale of the transformation function, add an additional intercept
parameter $-\alpha$, to arrive at the TM 
\begin{align*}
    (\pZ, (\bvec^\top, 1)^\top, (\hat\parm^\top, -\alpha)^\top)
\end{align*}
with distribution function
\begin{align*}
    F_{\rY\given\rx}(\ry \given \rx) = \pZ(\bvec(\ry,\rx)^\top\hat\parm - \alpha).
\end{align*}
Because $\hat\parm$ maximizes the likelihood, $\alpha$ is constrained to $0$.
The score residual for a single datum $y \in (\ubar{y}, \bar{y}]$ is now defined as
\begin{align} \label{eq:resid}
    r := \frac{\partial}{\partial\alpha} \ell(\parm, \alpha; \ry, \rx) 
        \bigg\rvert_{\hat\parm, \alpha \equiv 0},
\end{align}
which can be understood as the score contribution of a single observation
to test $\alpha = 0$ for a covariate which is not included in the model.
When viewed as a random variable, the vector of score residuals has mean zero
asymptotically and its components are asymptotically uncorrelated
\citep{farrington2000residuals}.
\end{definition}
The score residuals can be derived in closed form for a 
transformation model and observations under any form of uninformative
censoring
\begin{align}
r = \begin{cases}
   - \dZ'(\bvec(\ry,\rx)^\top\hat\parm)/\dZ(\bvec(\ry,\rx)^\top\hat\parm), & 
    \text{exact,} \\
   - \dZ(\bvec(\bar{\ry},\rx)^\top\hat\parm)/\pZ(\bvec(\bar{\ry},\rx)^\top\hat\parm), &
    \text{left,}\\
   \dZ(\bvec(\ubar{\ry},\rx)^\top\hat\parm)/(1 - \pZ(\bvec(\ubar{\ry},\rx)^\top\hat\parm)), & 
    \text{right,} \label{eq:explicitresiduals} \\
   (\dZ(\bvec(\ubar{\ry},\rx)^\top\hat\parm) - \dZ(\bvec(\bar{\ry},\rx)^\top\hat\parm) \big/ & 
        \text{interval.} \\
        \quad (\pZ(\bvec(\bar{\ry},\rx)^\top\hat\parm) - \pZ(\bvec(\ubar{\ry},\rx)^\top\hat\parm))
\end{cases}
\end{align}
\begin{example}[Lm, cont'd] \label{ex:resids}
By including the addtitional intercept parameter in the normal linear 
model in eq.~\eqref{eq:lm}, the score residuals are given by
\begin{align*}
  &\frac{\partial}{\partial \alpha} \ell(\eparm_1, \eparm_2, \shiftparm, \alpha; \ry, \rx) 
    \bigg\rvert_{\hat\eparm_1, \hat\eparm_2, \hat\shiftparm, \alpha \equiv 0} \\
    =\;&\frac{\partial}{\partial \alpha} \big\{ \log \phi\big(\eparm_1 + 
    \eparm_2 \ry - \rx^\top \shiftparm -\alpha) + \log(\eparm_2) \big\} \bigg\rvert_{\hat\eparm_1, \hat\eparm_2, \hat\shiftparm, \alpha \equiv 0} \\
  =\;&\hat\eparm_1 + \hat\eparm_2 \ry - \rx^\top \hat\shiftparm = \frac{y - \hat\beta_0 - \rx^\top \hat{\tilde\shiftparm}}{\hat\sigma}.
\end{align*}
In this simple case the score residuals are equivalent to scaled least-square residuals,
which underlines the more general nature of score residuals.
In Section~\ref{subsec:invariance} and Appendix~\ref{app:score-resid-forms}, we give further
examples and intuition on score residuals in non-linear and non-Gaussian settings.
\end{example}
We are now ready to cast transformation models into the framework of SEMs.
Here, it is natural to view the response $Y$ as a deterministic function
of the transformed random variable $\rZ \sim \pZ$, which is given by the
inverse transformation function $\h^{-1}$ in the following definition.
\begin{definition}[Structural equation transformation model]\label{def:tsem}
Let the conditional distribution of $\rY$ conditional on $\rX, \rH, \rA$ be given by the
transformation model $F_{\rY \given \rX, \rH, \rA} = \pZ \circ \h$.
The structural equation for the response is a deterministic function of $\rX$, $\rH$, $\rA$
and the exogenous $\rZ$, which, by definition, is distributed according to $\pZ$
and independent of $(\rX, \rH, \rA)$. Relationships other than the transformation
function are assumed to be linear. Taken together, the following SEM defines
a (partially) linear structural equation transformation model
\begin{align*}
    \rY &\leftarrow g(\rZ, \rX, \rH, \rA) := \h^{-1}(\rZ \given \rX, \rH, \rA) \\
    \rX & \leftarrow \mB_{\rX\rX} \rX + \mB_{\rX\rH} \rH + \mM_\rX \rA + \varepsilonvec_\rX \\
    \rH & \leftarrow \mB_{\rH\rH} \rH + \mM_\rH \rA + \varepsilonvec_\rH \\
    \rA &\leftarrow \varepsilonvec_\rA \\
    \rZ &\sim \pZ,
\end{align*}
where $\varepsilonvec_\rX, \varepsilonvec_\rH, \rA, \rZ$ are jointly independent.

As always, the structural equations are defined to hold as statements in distribution.
By Corollary~1 in \citet{hothorn2018most}, the transformation function $\h$ and its inverse
exist, are unique and monotone non-decreasing in $\rY$ and $\rZ$, respectively.
In contrast to the linear SEM in eq.~\eqref{eq:sem}, the SEM in Definition~\ref{def:tsem} is set up
involving a transformed response and a potentially non-linear inverse transformation $g$.
\end{definition}
However, from the perspective of transformation models it is more natural to
parameterize the transformation function $\h$ instead of its inverse, because
parameters in linear TMs are readily interpretable on this scale.
For the empirical evaluation of the proposed estimator, we set up the transformation function as
\begin{align} \label{eq:semtrafo}
   \h(\rY \given \rX, \rH, \rA) =
   \bvec(\ry)^\top\thetavec - \betavec^\top \rX - \mB_{\rY\rH}\rH 
   - \mM_{\rY}\rA.
\end{align}
A graphical representation of the SEM in Definition~\ref{def:tsem}
is shown in Figure~\ref{fig:tsem}.
The basis expansion $\bvec(\ry)^\top\thetavec$ in eq.~\eqref{eq:semtrafo} can be viewed 
as an intercept function, which fixes the overall shape of the transformation function. 
The remaining additive components of the transformation function,
in turn, solely shift the transformation up- or downwards with the covariates.
This may seem restrictive at first, however, all covariates influence not only
the conditional mean, but all higher conditional moments of $F_{\rY \given \rX, \rH, \rA}$.
We do not display the possibility that some components of $\rX$
directly influence each other, and likewise for $\rH$.
In fact, in the simulations in Section~\ref{sec:empirical}, 
the coefficients $\mB_{\rX\rX} = \mB_{\rH\rH} = 0$.

Next, we will present our main proposal on distributional anchor regression 
to achieve robust TMs with respect to perturbations on the anchor variables $\rA$.

\section{Distributional Anchor Regression} \label{sec:danchor}

We now formulate distributional anchor regression, for which we consider 
a distributional loss function, \ie the negative log-likelihood, which can take
into account censored observations and captures the conditional distribution
of $\rY \given \rX$, and a causal regularizer involving score residuals.
We first give some intuition how to arrive at the distributional anchor loss, 
starting from the linear $L_2$ anchor loss.
One can decompose the linear $L_2$ anchor loss
\begin{align*}
  L_2(\shiftparm; \yvec, \mX, \mA) =
  \norm{(\Id - \prm)(\yvec - \mX\shiftparm)}_2^2 +
   \gamma \norm{\prm(\yvec - \mX\shiftparm)}_2^2
\end{align*}
into a squared error and a pure penalty term.
We rewrite
\begin{align*}
  L_2(\shiftparm; \yvec, \mX, \mA) =
  \norm{\yvec - \mX\shiftparm}_2^2 + (\gamma - 1)
    \norm{\prm(\yvec - \mX\shiftparm)}_2^2,
\end{align*}
which is a sum of the squared-error loss and a causal regularizer involving the 
$L_2$ norm of the residuals $(\yvec - \mX\shiftparm)$ projected linearly onto 
the space spanned by the columns of $\mA$ to encourage uncorrelatedness between 
the residuals and the anchor variables. 
The cross-terms when expanding the $L_2$ norm vanish because $\prm$ is an 
orthogonal projection.
Now an intuitive choice for the distributional anchor regression loss would be
\begin{align*}
  L(\shiftparm; \yvec, \mX, \mA) 
  \propto - \sum_{i=1}^n \ell(\shiftparm; \ry_i, \rx_i) + (\gamma - 1) \norm{\prm\rvec}_2^2,
\end{align*}
where the negative log-likelihood induced by a transformation model, $\ell(\cdot)$, 
replaces the squared error loss and, most importantly, 
$\rvec$ denotes the vector of score residuals as defined in Section~\ref{sec:tram}.
Thus, the loss encourages uncorrelatedness between the anchor
variables and the score residuals, particularly for large values of $\gamma$.
The origin and importance of score residuals is outlined in Appendix~\ref{app:score-resids}.
We now give a definition for the distributional anchor loss.
\begin{definition}[Distributional anchor loss]
Consider a linear TM and its SEM, as in Definition~\ref{def:tsem}.
Then, the empirical distributional anchor loss is defined as 
\begin{align*}
L(\parm; \yvec, \mX, \mA, \xi) = 
    - \sum_{i=1}^{n} \ell(\parm; \ry_i, \rx_i)/n + \xi \norm{\prm \rvec}_2^2/n,
\end{align*}
where $\ell(\cdot)$ denotes the log-likelihood induced by a TM, $\rvec$ denotes 
the vector of score residuals and $\xi \in [0, +\infty)$ controls the extent 
of causal regularization.
As mentioned earlier, the log-likelihood is conditional on $\rX$.
\end{definition}
\begin{example}[Lm, cont'd]
For normal linear regression with constant variance, the linear $L_2$ anchor loss and the distributional
anchor loss are equivalent. This is because
\begin{align*}
    &L(\eparm_1, \eparm_2, \shiftparm; \yvec, \mX, \xi) \\
    = \; &- \sum_{i=1}^n \left\{ \log \phi(\eparm_1 + \eparm_2 \ry_i - \rx_i^\top\shiftparm) \right\}/ n -
       \log(\eparm_2)
       + \xi\nnorm{\prm(\eparm_1 + \eparm_2\yvec - \mX\shiftparm)}_2^2/n \\
    =\; &\nnorm{\yvec - \beta_0 - \mX\tilde\shiftparm}_2^2 / (2 \sigma^2 n) 
    + \xi\nnorm{\prm(\yvec - \beta_0 - \mX\tilde\shiftparm)}_2^2 / (\sigma^2 n) + C.
\end{align*}
Absorbing all additive constants into $C$ and factoring out the variance
renders the above objective equivalent to the linear $L_2$ anchor loss for $\xi = (\gamma - 1) / 2$.
\end{example}
Relaxing the assumptions of linear $L_2$ anchor regression towards censored outcomes and
non-linear expectation functions comes at the cost of some theoretical guarantees.
In the following, we discuss which aspects of the residual distribution remain invariant
in distributional anchor regression and issues in identification of the causal parameter when 
the anchors are valid instruments.

\subsection{Residual invariance} \label{subsec:invariance}

When considering the case $\xi \to \infty$, the solution to the distributional anchor
loss in the population version is a parameter $\parm^{\to \infty} = \lim_{\xi \to \infty} \parm(\xi)$,
which fulfills $\Corr(\rA, \rvec(\parm^{\to\infty})) = 0$ (assuming $\Ex[\rA] = 0$).
As in \citet{buehlmann2018invariance}, we consider the set
\begin{align}
    I = \{\parm \mid \Ex[\rA r(\parm)] = 0 \}, 
\end{align}
which contains solutions with zero correlation between anchors and score residuals. 
In particular, $\parm^{\to \infty} \in I$. 
We then arrive at a result similar to Proposition~5.1 in \citet{buehlmann2018invariance}.
\begin{proposition} \label{prop:residualinvariance}
Consider an anchor TM as in Def.~\ref{def:tsem} and assume $\Ex[\rA\rA^\top]$ to be positive definite.
Assume that $\Ex[r(\parm) \given \rA] = \mu_\parm + \zetavec_\parm^\top\rA$ is linear in $\rA$,
where $r(\cdot)$ denotes the score residual, $\mu_\parm \in \RR$, and $\zetavec_\parm \in \RR^{q}$.
Then, for any $\parm \in I$ we have for every $\pdo$-intervention on $\rA$ with $\pdo(\rA = \avec)$
\begin{align}
    \Ex[r(\parm) \given \pdo(\rA = \avec)] \equiv \mu_\parm,
\end{align}
which is independent of the (deterministic or random) value $\avec$.
\end{proposition}
Note that the linearity assumption
in the conditional expectation function of the residuals holds if 
the anchors are discrete \citep[][Corollary~5.1]{buehlmann2018invariance}.
The proof of Proposition~\ref{prop:residualinvariance} is analogous to the one given 
in \citet[Prop.~5.1]{buehlmann2018invariance}.
Proposition~\ref{prop:residualinvariance} shows that, similar to non-linear anchor
regression, the first moment of the score residuals conditional on the anchors
is invariant. However, it does not state any invariance properties for higher moments
or the entire residual distribution, which can be achieved in linear $L_2$ anchor regression
\citep[][Theorem~3]{rothenhaeusler2018anchor}. 

To build further intuition about residual invariance in distributional anchor regression,
the following example shows that score residuals are equivalent to martingale residuals 
in a parametric version of the Cox proportional hazard model 
\citep{barlow1988residuals,therneau1990martingale}. Further examples of 
score residuals for transformation models with probit and logit link are given in 
Appendix~\ref{app:score-resid-forms}.
\begin{example}[Cox regression]
The Cox proportional hazards (PH) model \citep{cox1975partial} is defined via
\begin{align*}
   \lambda(\ry\given\rx) = \lambda_0(\ry)\exp(\rx^\top\shiftparm),
\end{align*}
where $\lambda$ denotes the hazard function which is assumed to consist
of a baseline hazard $\lambda_0$ and a multiplicative contribution of the
covariates. The Cox PH model can be understood as a transformation model 
when choosing $\pZ(\rz) = 1 - \exp(-\exp(\rz))$. Then we have
\begin{align*}
    \h(\ry \given \rx) = \log\Lambda(\ry\given\rx) = \log\Lambda_0(\ry) + \rx^\top\shiftparm,
\end{align*}
where $\Lambda(\ry\given\rx) = \int_0^\ry\lambda(u\given\rx)\mathrm{d}u$ denotes the 
cumulative hazard and $\Lambda_0$ the cumulative baseline hazard function.
For a potentially right censored observation $(\ry, \rx, \delta)$, where 
$\delta \in \{0, 1\}$ denotes the event indicator (with 1 specifying an
exact and 0 a right censored observation), the likelihood contribution is
given by
\begin{align*}
    \ell(\parm,\shiftparm; \ry, \rx, \delta) = \delta\left[\h(\ry \given \rx) + 
        \log\h'(\ry\given\rx)\right] - \exp(\h(\ry\given\rx)).
\end{align*}
The score residual can be derived by including and restricting the additional
intercept $\alpha \equiv 0$ in the model in~\eqref{eq:resid} or via the general
form given in~\eqref{eq:explicitresiduals}
\begin{align*}
    r = \delta - \exp(\hat\h(\ry\given\rx)) = \delta - \hat\Lambda(\ry\given\rx), \; r \in (-\infty, 1].
\end{align*}
The above expression for the score residual is equivalent to martingale 
residuals, which quantify the discrepancy between the event indicator (``observed'')
and the estimated cumulative hazard (``predicted''), somewhat analogous to least squares residuals.
In contrast to least square residuals, martingale residuals have a skewed
distribution \citep[see, \eg][]{aalen2008survival}.

For distributional anchor regression, de-correlating the martingale residuals
from the anchors (\eg different countries or hospitals) stabilizes out-of-distribution
prediction in the sense that the first moment of the martingale residuals is invariant across 
intervention values of the anchors, due to Proposition~\ref{prop:residualinvariance}.
\end{example}

\subsection{Identifiability of the causal parameter}

Linear $L_2$ anchor regression is equivalent to instrumental variable regression
for $\gamma \to \infty$ (see Section~\ref{sec:bg}). 
As a consequence, the causal parameter is identified as long as the anchors are valid instruments
and the conditional expectation $\Ex[\rY \given \rX]$ is linear \citep{angrist1996identification}.
However, as soon as one leaves the linear regime, restrictive moment conditions have
to be employed, namely $\Ex[Y - f(\rX) \given \rA] = 0$. Then, $f$ is commonly fitted
using a generalized method of moments estimator (see, \eg \citet{Foster1997} for the logistic
linear case). 

The causal parameter is not identified in anchor TMs. 
The score-residual condition $\parm^{\to\infty} \in I$ is not directly related to a moment 
condition as described above. In fact, score residuals can
be interpreted as the slope of the negative log-likelihood contribution of a single
observation evaluated at the maximum-likelihood estimate (MLE). 
Thus, they also take into account the curvature of the log-likelihood 
(\ie the variance of the prediction instead of first moments only).
An empirical evaluation of the non-identifability is given in Section~\ref{sec:sim}.

We note that neither (non-) linear $L_2$ nor distributional anchor regression is targeted for estimating the
causal parameter. Instead, these methods aim to stabilize and improve worst-case prediction error
under perturbations.
Proposition~\ref{prop:residualinvariance} leads to an interpretation of stability: 
$\parm^{\to\infty}$ is the parameter for which the first conditional
moment of the residuals given the anchors remains invariant.

In the following Section, we will empirically
evaluate the prediction performance of transformation models estimated
under the distributional anchor loss. We empirically investigate non-identification 
of the causal parameter in two simulation scenarios in Section~\ref{sec:sim}.
Computational details for fitting TMs using the distributional anchor loss, 
are given in Appendix~\ref{app:comp}.

\section{Empirical Results} \label{sec:empirical}

We begin the section by describing the experimental setup in the application and
simulation studies and then present the results in Sections~\ref{sec:app}~and~\ref{sec:sim}.
We consider median housing prices in the BostonHousing2 dataset \citep{harrison1978hedonic} 
to illustrate an application of anchor TMs in normal linear regression (Lm),
which assumes normality and equal variances.
To lift these assumptions, a continuous outcome probit (c-probit) regression is used to
model more complex, skewed distributions, which are typical for housing prices.
Then, we use a continuous outcome logistic (c-logit exact) model, which enables more easily
interpretable shift effects on the log-odds scale. Lastly, we show the
c-logit (censored) model which now takes into account the censored observations in the 
BostonHousing2 dataset and retains interpretability of the parameters on the log-odds scale.
Furthermore, the proposed distributional framework for anchor regression is evaluated in
simulation studies for Lm, c-probit and ordered logistic regression (o-logit).
A summary of the models used to empirically evaluate anchor TMs
is given in Table~\ref{tab:models}.
\begin{table*}[!ht]
\centering
\caption{
Transformation models used to illustrate the distributional anchor
loss. $\pSL = \expit$ denotes the standard logistic distribution.
By $\tilde\yvec$ we denote the dummy encoded
response, $\tilde\yvec = \evec(k)$, for $\rY$ taking class $\ry_k$,
$k = 1, \dots, K$. Here, $\evec(k)$ denotes the $k$th unit vector.
In the experiments, the basis functions for $\ry$ are Bernstein polynomials
with maximum order $P$, $\bernx{P}(\ry) \in \RR^{P+1}$.
Because the transformation function $\h(\ry) = \bvec(\ry)^\top\parm$ must 
be monotone non-decreasing, we require some constraints on the parameters
of the transformation function.
}\label{tab:models}
\begin{tabular}{@{}llll@{}}
\toprule
\textbf{Name} & \textbf{Transformation Model} & \textbf{Constraints} &
\textbf{Type Response} \\ \midrule
Lm          
& 
$\left(\pN, (1, \ry, \rx^\top)^\top, 
(\theta_1, \theta_2, -\shiftparm^\top)^\top\right)$               
&
$\theta_2 > 0$
&
Continuous
\\
\\
c-probit
&
$\left(\pN, (\bernx{P}^\top, \rx^\top)^\top, 
(\thetavec^\top, -\shiftparm^\top)^\top\right)$               
&
$\theta_1 \leq \dots \leq \theta_{P+1}$
&
Continuous                       
\\
\\
c-logit
&
$\left(\pSL, (\bernx{P}^\top, \rx^\top)^\top, 
(\thetavec^\top, -\shiftparm^\top)^\top\right)$               
& 
$\theta_1 \leq \dots \leq \theta_{P+1}$
&
Continuous
\\
\\
o-logit
&
$\left(\pSL, (\tilde\yvec^\top, \rx^\top)^\top, 
(\thetavec^\top, -\shiftparm^\top)^\top\right)$               
&
$\theta_1 < \dots < \theta_{K - 1},$
&
Ordinal
\\
&
&
$\theta_K = + \infty$
&
\\
\bottomrule
\end{tabular}
\end{table*}

To evaluate prediction performance, we study the average negative log-likelihood (NLL). 
The NLL is a strictly proper scoring rule and thus encourages honest distributional 
forecasts \citep{good1952rational,gneiting2007strictly}.
In addition, the NLL is comparable across nested TMs and different choices for $\pZ$.

To study ``worst-case'' prediction performance, the quantile function of the individual
NLL contributions is studied.
Here, larger quantiles reflect harder-to-predict observations, since the model
assigns a low likelihood at the estimated parameters conditional on the observation. 
This approach is the distributional analog to studying quantiles of the squared errors 
\citep{rothenhaeusler2018anchor} or the absolute deviations \citep{buehlmann2018invariance}
in (non-) linear $L_2$ anchor regression.

\subsection{Application: BostonHousing2} \label{sec:app}

For the BostonHousing2 data we wish to predict corrected median housing prices (\code{cmedv})
from several socio-economical and environmental factors ($n = 506$).
These include per capita crime rates (\code{crim}), average number of rooms (\code{rm}),
and nitric oxide concentration (\code{nox}) among others.
Each observation corresponds to a single census tract in Boston.
Individual cities will serve as anchors in this example because they are
plausibly exogenous factors that induce heterogeneity in the observed covariates
and housing prices.

\begin{figure*}[!ht]
\centering
\includegraphics[width=0.9\textwidth]{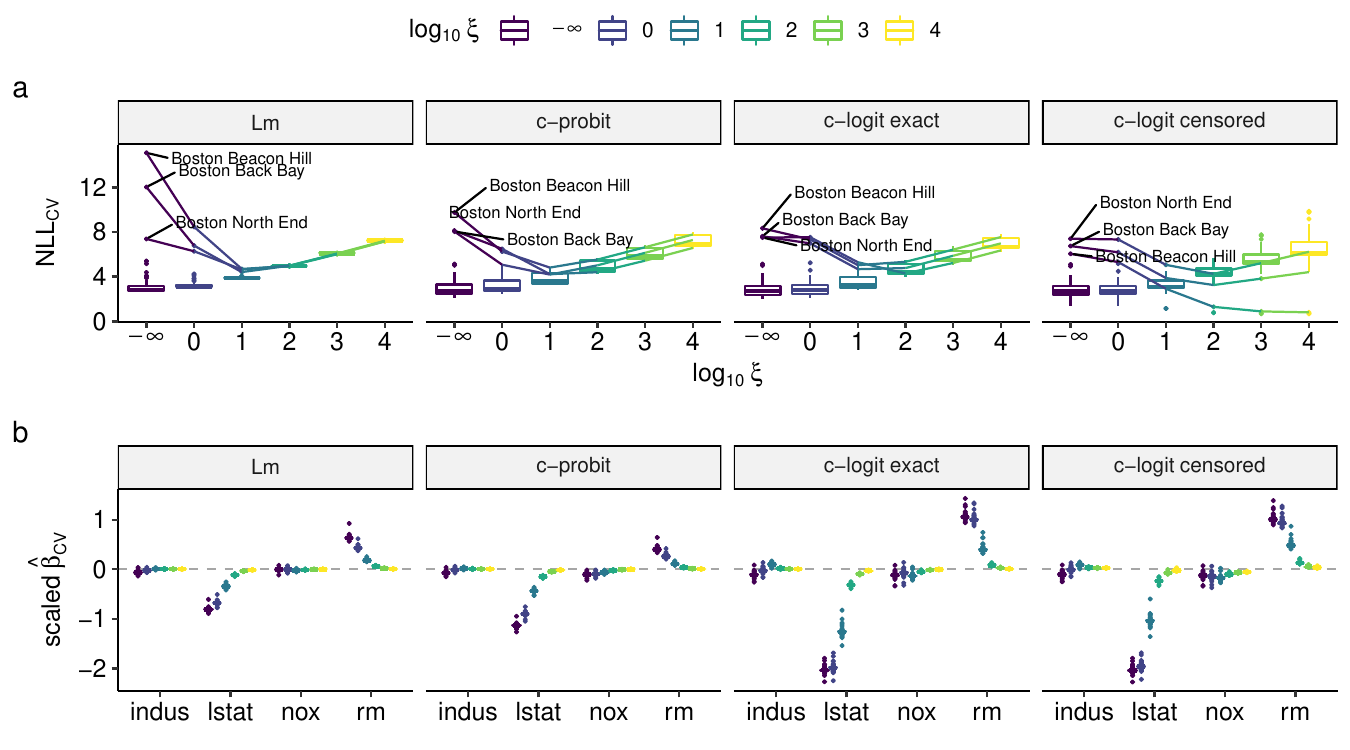}
\caption{
Leave-one-environment-out cross validation under increasing causal regularization 
for the BostonHousing2 data, with \code{town} as anchors.
A linear (Lm), continuous probit (c-probit) and continuous logit (c-logit, using the
exact and censored response) model is fitted on 91 towns and used to predict the left out town. 
\textsf{a}: Mean out-of-sample NLL for the left-out census tracts.
Beacon Hill, Back Bay and North End are consistently hardest to predict.
Consequently, for these towns the cross-validated NLL improves with increasing
causal regularization up to a certain point.
For the majority of the remaining towns prediction performance decreases. 
We thus indeed improve worst-case prediction, in analogy to eq.~\eqref{eq:anchorpop}.
Note that $\log_{10} \xi = - \infty$ corresponds to the unpenalized model.
\textsf{b}: Scaled regression coefficients, 
which are interpretable as difference in means (Lm), difference
in transformed means (c-probit) and log odds-ratios (c-logit) per standard
deviation increase in a covariate.
Solely the c-logit (censored) model accounts for right-censored observations.
With increasing causal regularization the estimates shrink towards zero,
indicating that \code{town} may be a weak instrument (see Appendix~\ref{app:invalid}).
} \label{fig:res:bh2}
\end{figure*}

``Leave-one-environment-out'' (LOEO) cross validation is used to demonstrate the
change in estimated regression coefficients and NLL comparing a plain model
without causal regularization ($\xi = 0$) to three different anchor TMs over
a large range of causal regularization (Figure~\ref{fig:res:bh2}).
For some of the left-out towns the conditional distribution of \code{cmedv}
will differ from the training distribution and contain unseen perturbations.
In this case, housing prices of this town will be harder to predict, leading 
to a worse cross-validated NLL compared to the environments which are not perturbed.
We hypothesize, an anchor TM should improve prediction performance for the census
tracts in these hard-to-predict towns, in analogy to the distributional robustness 
results implied by eq.~\eqref{eq:anchorpop}, whereas it should perform worse than a 
plain TM for environments with only mild perturbations.

First, a linear model assuming homoscedasticity and conditional normality
is used to estimate the conditional distribution of \code{cmdev} depending
on the socio-economic factors described above.
A notable reduction in the observed worst-case loss is already observed for
mild causal regularization ($\xi \in \{1, 10\}$) without losing too much predictive
performance for the other environments (Figure~\ref{fig:res:bh2}\textsf{a}).
For stronger causal regularization, the mean cross-validated NLL becomes gradually worse.
However, assuming a symmetric distribution for prices ignores the typical
skewness of these outcomes.
The c-probit model estimates a non-linear basis expansion in the response and
thus relaxes the homoscedasticity and conditional normality assumption.
When switching from $\pZ = \Phi$ to $\pZ = \expit$, the interpretation
of $\h$ changes from a latent normal to a logistic scale. The former has no direct
interpretation, whereas for the latter, $\shiftparm$ can be interpreted as log odds-ratios.
A similar gain in terms of worst-case CV NLL is observed for the c-probit
model compared to Lm.

\begin{figure}[!ht]
\centering
\includegraphics[width=0.7\textwidth]{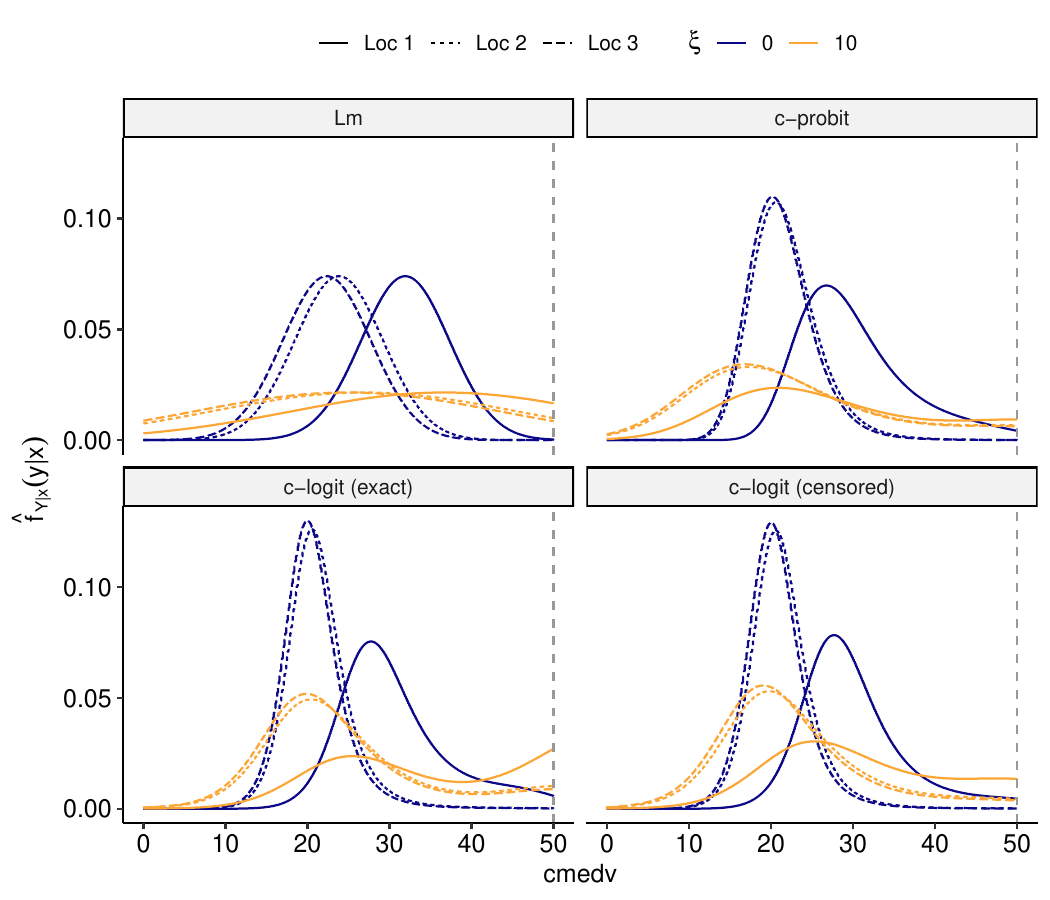}
\caption{
Density estimates for the three census tracts (\code{Loc~1}, \code{Loc~2}, \code{Loc~3}) 
in Boston Beacon Hill, the hardest to predict town in terms of LOEO cross-validated NLL
for $\xi = 10$ (\cf Figure~\ref{fig:res:bh2}). The dashed gray line indicates the 
observed outcomes for all three locations, which were all censored at $\$50'000$.
Lm assumes equal variances and conditional normality, 
whereas c-probit loosens this assumption leading to more accurate, skewed distributions.
Only c-logit (censored) takes into account right censoring in the data and puts a smaller
probability density on $\$50'000$ than the c-logit (exact) model, which ignores the 
censoring. 
} \label{fig:res:bh2d}
\end{figure}

Figure~\ref{fig:res:bh2d} shows the predicted conditional densities for the
three observations in Boston Beacon Hill and emphasizes the importance of
modelling \code{cmedv} using a right-skewed distribution. 
The densities are shown for the regularized transformation model ($\xi = 10$)
and the unregularized model ($\xi = 0$). For all regularized models, a flatter 
(\ie more uncertain) distribution is predicted, putting more probability mass
on the values beyond $\$50'000$.

A disadvantage of switching from a linear probit to a non-linear probit model
is the loss of interpretability of the individual regression coefficients
\citep[\eg][]{fahrmeir2007regression}.
However, also this disadvantage can be overcome in the framework of anchor TMs
by specifying a different inverse link function, $\pZ$, while keeping the
basis expansion in the outcome equally flexible.
The c-logit (exact) model allows the interpretation of $\hat\shiftparm$ on
the log-odds scale and shows a similar gain in worst-case prediction to c-probit.
However, the housing prices above $\$50'000$ (\code{cmedv} = 50) are 
right-censored in the BostonHousing2 data, which is commonly ignored in analyses,
but crucial to capture the uncertainty in predicting the skewed outcome
\citep{gilley1996harrison}.
The c-logit (censored) model now takes into account right censoring of the observations
and still yields regression coefficients that are interpretable as log odds-ratios
\citep{lohse2017continuous}.
Indeed, for census tract \code{Loc~1} the censored c-logit model reflects the higher
uncertainty implied by the censored responses compared to the c-probit or the exact
c-logit model (Figure~\ref{fig:res:bh2d}). Note how c-logit (exact) attributes high
density to \code{cmedv} $=50$ by mistakenly treating the censored observations as exact,
because the likelihood contribution is the density instead of the area under the density
to the right of the censored observation.

Taking into account right censoring apparently facilitated out-of-sample prediction for 
Boston Beacon Hill, Back Bay and North End, but the improvement through
causal regularization diminished slightly compared to c-probit or Lm 
(Figure~\ref{fig:res:bh2}\textsf{a}).

Scaled coefficient estimates for all three models are shown in Figure~\ref{fig:res:bh2}\textsf{b}.
With increasing amount of causal regularization, all estimates shrink towards 0,
which indicates \code{town} may be a weak instrument \citep{imbens2005robust}, 
for more details see Appendix~\ref{app:invalid}.
However, intermediate amounts of causal regularization yield estimates for which
anchors and score residuals are somewhat de-correlated and still lead to the desired
robustness against perturbations in unseen environments.

\subsection{Simulations} \label{sec:sim}

In this section, the results of the simulation scenarios are presented. 
The scenarios along with parameters for the SEMs used to simulate from the models 
in Table~\ref{tab:models} are summarized in Table~\ref{tab:sim}.

\begin{table*}[!ht]
\centering
\caption{%
Simulation scenarios used to empirically evaluate distributional
anchor regression.
Scenarios {\sc la} and {\sc nla} are adapted from \citet{buehlmann2018invariance}
and will be used to evaluate linear and continuous probit anchor TMs.
The eight covariates omitted in the table in both scenarios are noise covariates, \ie
$\beta_j = 0, \; j \neq 2,3$.
In {\sc nla}, $f(\rX) = X_2 + X_3 + \I(X_2 \leq 0) + \I(X_2 \leq -0.5)\I(X_3 \leq 1)$.
Both use $n_{\text{train}} = 300$ and $n_{\text{test}} = 2000$ observations.
In the {\sc iv} scenarios the instrumental variables assumptions
hold, because the anchor neither influences the hidden confounders nor the response.
The scenarios generalize Example~2 in \citet{rothenhaeusler2018anchor} to
anchor TMs with a continuous outcome ({\sc iv1}) and an ordinal outcome ({\sc iv2}).
Both use $n_{\text{train}} = 1000$ and $n_{\text{test}} = 2000$ observations.
A schematic DAG is shown together with the equations for all endogenous variables 
underneath. 
Note that for scenario \textsc{nla} which was taken from \citet{buehlmann2018invariance},
the data are not generated from the c-probit model but via the structural equation given 
below the DAG. For scenario \textsc{iv1}, in contrast, the data are generated via the 
c-probit model using the order 6 Bernstein polynomial basis (\ie the best approximation 
to $\h$ is used).
}\label{tab:sim}
\resizebox{0.85\textwidth}{!}{%
\begin{tabular}{@{}lllm{4cm}l@{}}
\toprule
\textbf{Scenario} & \textbf{Model} &
\textbf{Distributions}  & \textbf{Graphical model} & \textbf{Test perturbation} \\
\midrule
{\sc{la}}
&
Lm
&
$\begin{array}{rl}
    \veY &\sim \calN(0, 0.25^2) \\
    \veX &\sim \calN_{10}(0, \Id) \\
    \ve_H &\sim \calN(0, 1) \\
    \veA &\sim \calN_2(0, \Id) \\
    \rM_{\rX} &\sim \calN_2(0, \Id) \\
\end{array}$
&
\begin{tikzpicture}
 \node (X) {$\rX$};
 \node [above= of X] (E) {$\rA$};
 \node [right=2cm of X] (Y) {$\rY$};
 \node [above =of Y] (H) {$H$};
 \draw[Arrow] (E) -- (X);
 \draw[Arrow] (X) -- (Y);
 \draw[Arrow] (H) -- (X);
 \draw[Arrow] (H) -- (Y);
 \draw[Arrow] (E) -- (Y);   
\end{tikzpicture}
&
$\rA \sim \calN_2(0, 10\Id))$ \\
\cline{3-5} \\[-5pt]
&&\multicolumn{3}{c}{$Y = 3X_2 + 3X_3 + H - 2A_1 + \varepsilon_Y$} \\
&&\multicolumn{3}{c}{$X_j = H + \mM_{\rX}\rA + \varepsilon_{X_j}, \; j = 1, \dots, 10$} \\
\\[-5pt] \midrule
{\sc{nla}}
&
c-probit
&
$\begin{array}{rl}
    \veY &\sim \calN(0, 0.25^2) \\
    \veX &\sim \calN_{10}(0, 0.5^2 \Id) \\
    \ve_H &\sim \calN(0, 1) \\
    \veA &\sim \calN_2(0, \Id)
\end{array}$
&
\begin{tikzpicture}
 \node (X) {$\rX$};
 \node [above= of X] (E) {$\rA$};
 \node [right=2cm of X] (Y) {$\rY$};
 \node [above =of Y] (H) {$H$};
 \draw[Arrow] (E) -- (X);
 \draw[Arrow] (X) -- (Y);
 \draw[Arrow] (H) -- (X);
 \draw[Arrow] (H) -- (Y);
\end{tikzpicture}
&
$\begin{array}{l}
    \rA \sim \calN_2(\muvec, \Id))\\
    \muvec \sim \calN_2(10, 2\Id)
\end{array}$ \\
\cline{3-5} \\[-5pt]
&&\multicolumn{3}{c}{$Y = f(X_2, X_3) + 3H + \varepsilon_Y$} \\
&&\multicolumn{3}{c}{$X_j = 2H + A_1 + A_2 + \varepsilon_{X_j}, \; j = 1, \dots, 10$} \\
\\[-5pt] \midrule
{\sc{iv1}}
&
c-probit
&
$\begin{array}{rl}
    \ve_X &\sim \calN(0, 0.75^2) \\
    \ve_H &\sim \calN(0, 0.75^2) \\
    \ve_A &\sim \Rade \\
    \h &= \Phi^{-1} \circ F_{\chi^2_3}\\
\end{array}$
&
\resizebox{\linewidth}{!}{%
\begin{tikzpicture}
 \node (X) {$X$};
 \node [above= of X] (E) {$A$};
 \node[circle,draw] [right=2cm of X] (Y) {$Y$};
 \node [above =of Y] (H) {$H$};
 \node [right =of Y] (Z) {$Z$};
 \draw[Arrow] (E) -- (X);
 \draw[Arrow] (X) -- (Y);
 \draw[Arrow] (H) -- (X);
 \draw[Arrow] (H) -- (Y);
 \draw[Arrow] (Z) -- (Y);
\end{tikzpicture}}
&
$\begin{array}{l}
    \pdo(A=3.6)
\end{array}$ \\
\cline{3-5} \\[-5pt]
&&\multicolumn{3}{c}{$\h(\rY\given X, H, A) = \bernx{6}(\rY)^\top\thetavec + 0.3X + 0.6H$} \\
&&\multicolumn{3}{c}{$X = 0.6H + 0.3A + \varepsilon_{X}$} \\
\\[-5pt] \midrule
{\sc{iv2}}
&
o-logit
&
$\begin{array}{rl}
    \ve_X &\sim \calN(0, 0.75^2) \\
    \ve_H &\sim \calN(0, 0.75^2) \\
    \ve_A &\sim \Rade \\
    \h &= \pSL^{-1} \circ \Id \\
    K &\in \{4, 6, 10\} \\
\end{array}$
&
\resizebox{\linewidth}{!}{%
\begin{tikzpicture}
 \node (X) {$X$};
 \node [above= of X] (E) {$A$};
 \node[circle,draw] [right=2cm of X] (Y) {$Y$};
 \node [above =of Y] (H) {$H$};
 \node [right =of Y] (Z) {$Z$};
 \draw[Arrow] (E) -- (X);
 \draw[Arrow] (X) -- (Y);
 \draw[Arrow] (H) -- (X);
 \draw[Arrow] (H) -- (Y);
 \draw[Arrow] (Z) -- (Y);
\end{tikzpicture}}
&
$\begin{array}{l}
    \pdo(A=a) \\
    a \in \{1, 1.8, 3\}
\end{array}$ \\
\cline{3-5} \\[-5pt]
&&\multicolumn{3}{c}{$\h(\ry_k \given X, H, A) = \theta_k + 0.5X + 1.5H, \; k = 1, \dots, K - 1$} \\
&&\multicolumn{3}{c}{$X = 1.5H + M_{X}A + \varepsilon_{X}, \; M_X \in \{0.5, 1, 2\}$} \\
\\[-5pt] 
\bottomrule
\end{tabular}
}
\end{table*}

\subsubsection{Experimental Setup}

We begin with a comparison of linear $L_2$ anchor regression and the distributional
version of linear anchor regression in scenario {\sc la}, which was first used
to study anchor regression in \citet{buehlmann2018invariance}.
The non-linear scenario {\sc nla} also stems from \citet{buehlmann2018invariance},
which we use to show how shift transformation models can estimate non-linear
conditional expectation function, albeit for their linear model formulation in the
covariates.
For the last two scenarios {\sc iv1} and {\sc iv2}, the IV assumptions hold,
\ie the anchors influence only $\rX$.
Scenario {\sc iv1} showcases discrete anchors and a continuous response and a
non-linear transformation function, which we model by a continuous probit regression.
Scenario {\sc iv2} features an ordinal response and a more detailed simulation,
including various shift strengths.
In scenarios {\sc la}, {\sc iv1} and {\sc iv2}, the effect from $\rX$ to $\rY$
is denoted by $\shiftparm$, whereas the non-linear $f$ is used in scenario {\sc nla}.
For the data generating processes that involve transformation models,
the transformation function $\h$ is specified. 
For ordinal responses the number of classes, $K$, and for continuous outcomes, 
the maximum order of the Bernstein basis, $P$, determines the number of parameters
for the baseline transformation.
The parameters of the Bernstein basis are fixed by applying the transformation
function $\h$ to a $(P+1)$-vector of evenly spaced values in the desired support of $\rY$. 
In turn, such a basis approximation leads to a distribution approximation for the true
distribution of $\rY$ which improves as $P$ increases.
However, the transformation function is constrained to be monotone non-decreasing,
which makes a parsimonious parametrization sufficient.

\subsubsection{Scenario {\sc la}} \label{sim:la}

\begin{figure}[!ht]
\centering
\includegraphics[width=0.7\textwidth]{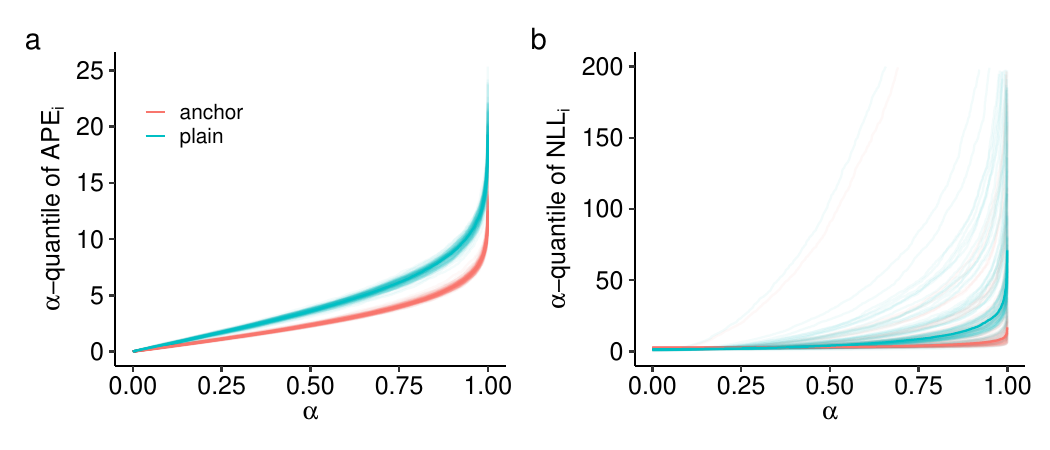}
\caption{
Test performance (thin lines) over 100 simulations for scenario {\sc la} 
with $n_{\text{train}} = 300$ and $n_{\text{test}} = 2000$. Median test
performance over all simulations is indicated by the thick line.
The $\alpha$-quantiles of test absolute prediction error 
$\text{APE} := \lvert \ry - \hat \ry \rvert$, where $\hat{\ry}$ denotes the 
conditional median, is shown for linear $L_2$ anchor regression (\textsf{a})
using $\gamma = 13$ and the negative log-likelihood contributions for distributional 
(conditionally Gaussian) linear anchor regression (\textsf{b}) with
$\xi = (\gamma - 1) / 2 = 6$.
The two models are equivalent up to estimating the residual variance via 
maximum likelihood in the distributional anchor TM.
The change in perspective from an $L_2$ to a distributional loss requires different
evaluation metrics, of which the log-likelihood, being a proper scoring rule, 
is the most natural choice.
} \label{fig:res:la}
\end{figure}

The linear anchor scenario {\sc la} was first presented in \citet{buehlmann2018invariance}
for the linear $L_2$ anchor loss.
The performance gain of using anchor regression compared to a plain linear
model is shown in Figure~\ref{fig:res:la} for the linear $L_2$ anchor loss (\textsf{a})
and the distributional anchor loss (\textsf{b}).

A performance gain across all quantiles of the log-likelihood contributions 
can be observed. However, the larger the quantile, the higher the performance gain.
The extent of causal regularization was chosen based on the theoretical
insight that, in a multivariate normal model, $\gamma$ can be interpreted as the
quantile of a $\chi^2_1$ distribution, which relates the expected size of the
unobserved perturbations to the conditional mean squared error given the anchors
\citep[Lemma~1 in][]{rothenhaeusler2018anchor}.
The variability in the NLL's quantile function in Figure~\ref{fig:res:la}\textsf{b}
appears to be larger than for the absolute prediction error (in panel \textsf{a}).
We attribute this to the sensitivity of the NLL towards worst-case prediction errors,
\ie for likelihood contributions close to zero, the NLL quickly diverges to
minus infinity. The APE as a measure of central tendency is not as sensitive to
worst-case prediction errors.

\subsubsection{Scenario {\sc nla}} \label{sim:nla}

\begin{figure}[!ht]
\centering
\includegraphics[width=0.7\textwidth]{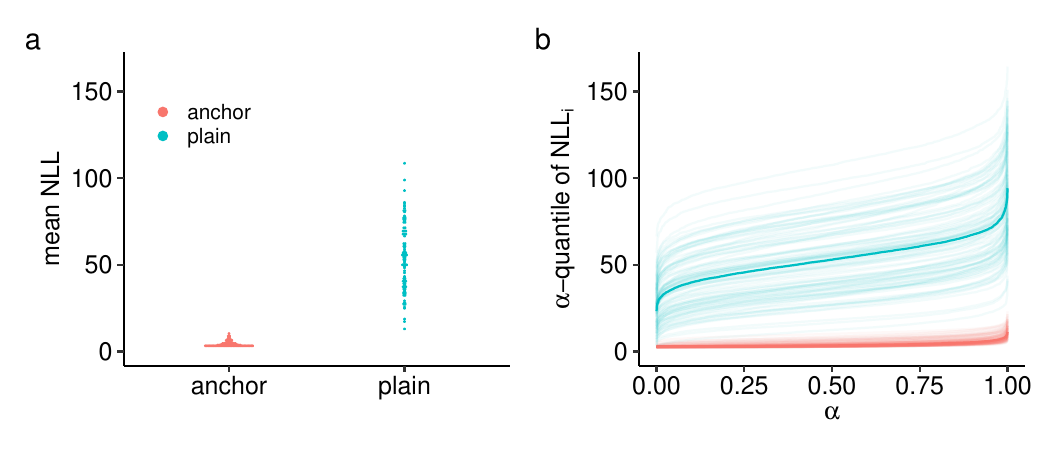}
\caption{
Test performance over 100 simulations for scenario {\sc nla} with 
$n_{\text{train}} = 300$ and $n_{\text{test}} = 2000$.
Mean (\textsf{a}) and $\alpha$-quantiles of the negative 
log-likelihood contributions (\textsf{b}) for the c-probit anchor TM.
The test data are generated under strong push-interventions on the distribution
of the anchors (\cf Table~\ref{tab:sim}). The strength of causal regularization
was chosen as $\xi = 6$.
} \label{fig:res:nla}
\end{figure}

In scenario {\sc nla}, which features non-linear anchor regression, 
a continuous probit model is fitted.
Figure~\ref{fig:res:nla} shows a vast gain in performance over all quantiles
of the NLL, comparable to what was observed in \citet{buehlmann2018invariance} 
with $L_2$ anchor boosting for quantiles of the absolute prediction error.
This gain in performance can be explained in the causal generalization framework
of \citet{christiansen2020causal}, because the causal function linearly
extends outside the support of $\rX_{\text{train}}$.
Note that although the graphical model representation suggests that the assumptions of IV 
regression hold, the conditional expectation is non-linear.
Additional simulations for a mis-specified normal linear anchor TM are given in
Appendix~\ref{app:sim}, to warrant the use of a non-linear model.

In some applications, a point prediction may be more desirable than a distributional
forecast. Anchor TMs can produce point estimates via the conditional quantile function,
\eg the conditional median. However, in these settings we recommend (non-) linear
$L_2$ anchor regression which is tailored specifically towards conditional mean estimation.
In Appendix~\ref{app:sim}, we present additional results for scenario {\sc nla} to
compare the performance of these point estimates from anchor TMs with results from $L_2$ anchor
boosting \citep{buehlmann2018invariance} using a combined linear model, random forest
base learner for reference. In essence, the conditional median predictions from anchor TM show a 
similar gain in worst-case absolute prediction error and can compete with the conditional 
mean predictions obtained from $L_2$ anchor boosting in this scenario.

\subsubsection{Scenario {\sc iv1}} \label{sim:iv1}

In scenario {\sc la} the anchors influence the response,
violating the instrumental variable assumptions.
Scenario {\sc iv1} explores binary anchors as valid instruments,
while the baseline transformation $\bernx{6}(\ry)^\top \thetavec$
is non-linear.

Note that although the model formulation is linear in $\shiftparm$, the conditional
expectation function may be non-linear, because of the non-linear
transformation function. 
This scenario is inspired by Example~2 in \citet{rothenhaeusler2018anchor}
and translates it into a transformation model SEM from Definition~\ref{def:tsem}
for continuous but non-normal outcomes.
\begin{figure}[!ht]
\centering
\includegraphics[width=0.7\textwidth]{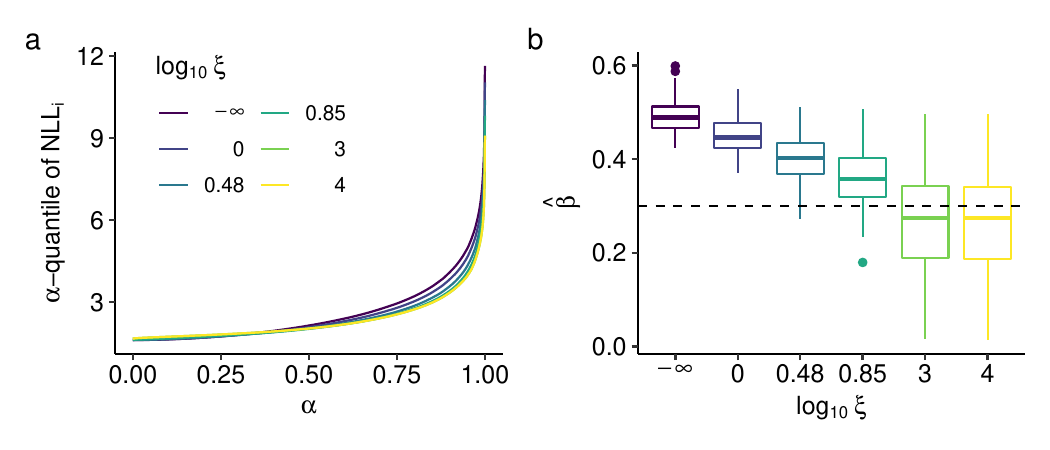}
\caption{
Test performance over 100 simulations for scenario {\sc iv1} with 
$n_{\text{train}} = 1000$ and $n_{\text{test}}=2000$.
Quantiles of the individual negative log-likelihood contributions (\textsf{a}) and estimates
of $\beta$ (\textsf{b}) for increasingly strong causal regularization.
The ground truth is indicated by a dashed line.
The test data are generated under the intervention $\pdo(\rA = 3.6)$.
} \label{fig:res:iv1}
\end{figure}
The test data were generated using $\pdo(A=3.6)$, for which a better predictive
performance under stronger causal regularization was observed (Figure~\ref{fig:res:iv1}\textsf{a}).
Additionally, although $\rA$ is a valid instrument, the causal parameter is
biased for larger $\xi$ (Figure~\ref{fig:res:iv1}\textsf{b}), due to the non-linear
conditional expectation $\Ex[\rY\given X]$.
Additional simulations for a mis-specified normal linear anchor TM fitted to data
generated under scenario {\sc iv1} are given in Appendix~\ref{app:sim}.

\subsubsection{Scenario {\sc iv2}} \label{sim:iv2}

The instrumental variable assumptions hold also in the last scenario {\sc iv2}.
However, the response's distribution is now ordered categorical and more varying
parameters are considered, including the number of classes of the response, 
the strength of the instruments and the perturbations in the test data (\cf Table~\ref{tab:sim}).
Note that the ordinal response may be viewed as an interval censored version of an 
underlying latent continuous response with common censoring points. This point of
view is especially useful for sampling from such a model and understanding it as a
transformation model.
\begin{figure*}[!ht]
\centering
\includegraphics[width=0.9\textwidth]{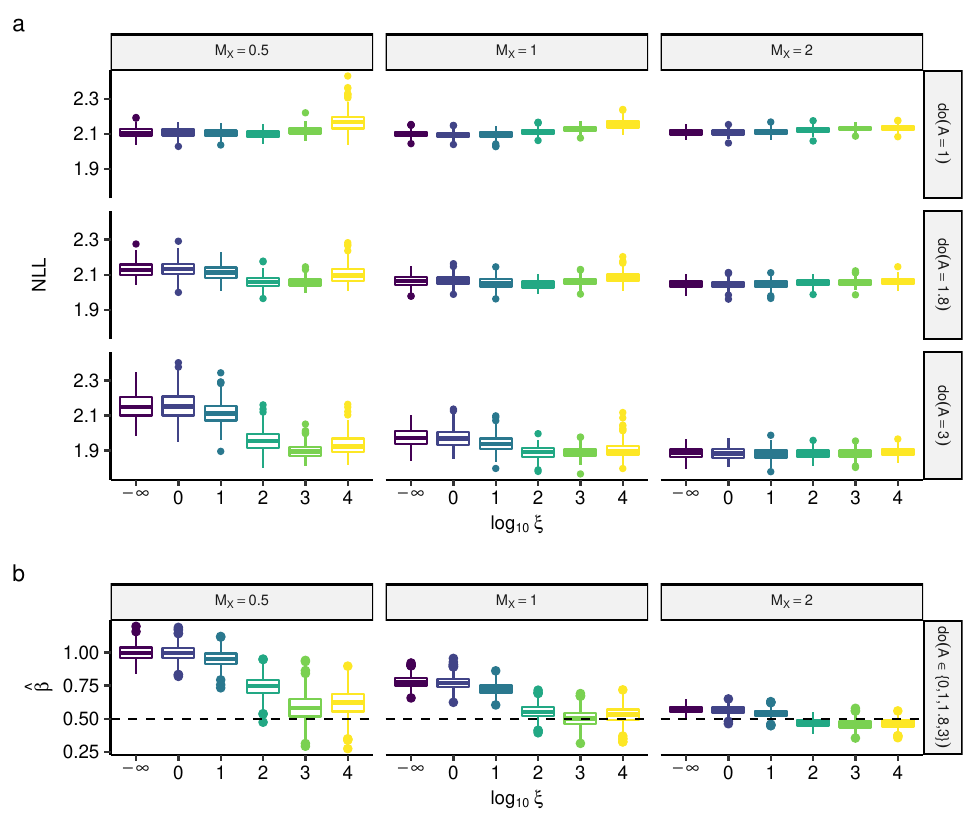}
\caption{
Test performance and coefficient estimates over 200 simulations for scenario {\sc iv2}.
Because the results are comparable for differing sample sizes and numbers of
classes, solely the results for $n_{\text{train}} = 1000$ and $K = 10$ are displayed.
\textsf{a}: Test log-likelihood contributions for varyingly strong instruments 
(columns) and perturbation sizes (rows).
\textsf{b}: Parameter estimates $\hat\beta$ for all
intervention scenarios together, since they do not influence estimation.
The simulated ground truth $\beta = 0.5$ is indicated with a dashed line.
} \label{fig:res:iv2}
\end{figure*}

Figure~\ref{fig:res:iv2} depicts test NLL alongside the estimated shift parameter, 
$\hat\beta$.
Also here, in case of strong perturbations anchor regression protects against unseen 
perturbations for larger $\xi$ (\eg $\pdo(A = 1.8)$ and $\pdo(A = 3)$ for $\mathrm{M}_{X} = 0.5$)
resulting in improved test predictions.
However, if the shift perturbations are not innovative, test prediction suffers with
increasing amounts of causal regularization (\eg $\pdo(A = 1)$ for $\mathrm{M}_{X} = 2$).
Note the interplay between the strength of the anchors, $\mathrm{M}_X$, and the
strength of the shift interventions. 
For larger $\mathrm{M}_{X}$, the training data becomes more heterogeneous and the larger
shifts are not as innovative, resulting in a weaker performance of anchor TMs for
increasing $\xi$.
Again, the estimated shift parameter is biased (Figure~\ref{fig:res:iv2}\textsf{b}).

The additional simulation results for $K = 4$ and $K = 6$ are shown in Appendix~\ref{app:sim}.
The conclusions are unaffected at those values for $K$. However, having a bounded
sample space for the response can be problematic, namely if shift perturbations are extremely
large and the model is linear in $\rA$. Then, it may happen that the response's marginal
distribution is extremely skewed towards class $1$ or $K$. Most prominently, this problem
appears in the binary case.

\section{Discussion and Outlook} \label{sec:discussion}

The proposed method of distributional anchor regression generalizes 
(non-) linear anchor regression beyond the assumptions of normality and 
homoscedasticity and beyond estimating solely a conditional mean.

In an exemplary analysis of the BostonHousing2 data we have illustrated
the flexibility of anchor TMs and demonstrated a gain in prediction performance
in terms of worst-case cross validated log-likelihood, while preserving interpretability
and appropriately accounting for censored observations.
The simulations show comparable results to established linear and non-linear
anchor regression models under both IV and invalid IV scenarios and 
extend the notion of invariance between residuals and environments to other 
than continuous responses.
Although anchor TMs are generally unable to recover the causal parameter, 
we argue that the ``diluted causal'' \citep{buehlmann2018invariance} parameter,
$\hat\bv^{\to\infty} := \hat\shiftparm(\xi)$ as $\xi \to \infty$, 
is interesting in its own right, 
for it induces invariance between anchors and the first conditional moment of the score residuals. 
In this sense, it allows robust test predictions in the presence of distributional shifts.
Much like the causal parameter, the diluted causal parameter leads to (aspects of)
invariant conditional distributions across environments.
Even though the powerful causal interpretation is lost, distributional anchor
regression yields models that allow causally flavored interpretations in terms of
stabilization and robustification across environments.

Anchor TMs estimate the whole conditional distribution and thus
enable robust prediction of a multitude of responses, which we
demonstrated for (censored) continuous and ordered categorical responses.
Possible extensions of anchor TMs are numerous. 
For instance, other types of responses include count and time-to-event data.
The framework of anchor TMs contains a fully parametric version of the Cox 
proportional hazard model \citep{hothorn2018most}, although an extension to 
classical survival models is also possible.
For instance, the Cox proportional hazard model \citep{cox1972regression} 
can be fitted by substituting the likelihood for the partial likelihood 
\citep{cox1975partial} in the distributional anchor loss, 
while the score residuals are equivalent to martingale residuals
\citep[\cf Appendix~\ref{app:score-resids};][]{barlow1988residuals,therneau1990martingale}.
As in high-dimensional linear and non-linear anchor regression,
anchor TMs could be fitted under a lasso penalty \citep{tibshirani1996regression}.
The idea of using a different class of residuals can also be translated
to other model classes, such as deviance residuals for GLMs, as long as
the theoretical requirements discussed in Section~\ref{sec:danchor} are met.

In terms of future work, further theoretical investigation of the distributional 
anchor loss, such as bounds on the generalization error, is warranted.
So far we restricted distributional regression to linear (in $\rx$) TMs because of their 
already highly flexible nature and simplicity in the considered DGPs.
However, more complex experimental designs require, for instance, random effects
or time-varying effects of covariates in time-to-event data.
Taken together, anchor TMs lay the foundation for future work on distributional
extensions of anchor regression.


\paragraph{Acknowledgements}
We thank Torsten Hothorn for fruitful discussions and input on the exemplary application.
We thank Malgorzata Roos for her helpful comments on the manuscript. Furthermore, we are
grateful all anonymous reviewers for their valuable feedback which improved both the 
theoretical contributions and the empirical presentation of our method.
The research of L. Kook and B. Sick was supported by the Novartis Research Foundation 
(FreeNovation~2019).
The research of P. B\"{u}hlmann was supported
by the European Research Council (ERC) under the European Union’s Horizon 2020 research 
and innovation programme (grant agreement No. 786461).


\vskip 0.2in
\bibliographystyle{plainnat}
\bibliography{bibliography}


\appendix
\renewcommand{\thesection}{\Alph{section}}
\counterwithin{figure}{section}
\renewcommand\thefigure{\thesection\arabic{figure}}
\counterwithin{table}{section}
\renewcommand\thetable{\thesection\arabic{table}}
\section{Notation} \label{app:notation}

Random variables are written in uppercase italic, $\rY$, and realizations 
thereof in lowercase, $\ry$. 
When stacked to a vector of $n$ observations, we write 
$\yvec = (\ry_1, \dots, \ry_n)^\top$.
Random vectors are written like random variables, but in bold, $\rX$, and
realizations thereof in lowercase bold, $\rx$. 
Stacked to a matrix for $n$ observations, 
we write $\mX = (\rx_1^\top, \dots, \rx_n^\top)^\top \in \RR^{n\times p}$.
Matrices are written in bold, normal uppercase, $\mA$, vectors
in bold italic lowercase, $\avec$.
The probability measure of a random variable $\rY$ is denoted by $\Prob_\rY$.
Coefficient matrices in the SEMs are denoted by $\mM$ for the anchors $\rA$
and by $\mB$ for all other components.
When restricting the coefficient matrix $\mB$ to a single component, 
\eg the effect of $\rX$ on $\rY$, we write $\mB_{\rY\rX}$.
Because for $\mM$ it is clear from context, we omit the $\rA$ in the index.

\section{Background on Score Residuals} \label{app:score-resids}

\citet{stigler2016seven} considers residuals the seventh
``pillar of statistical wisdom'', which highlights their conceptual importance.
Here, we will briefly justify the use of the score residual
in transformation models based on the work of \citet{lagakos1981residuals}, who
introduced martingale residuals for survival analysis under interval
censoring. \citet{farrington2000residuals} presents a 
summary of the different types of residuals used in survival analysis.
The scope of score residuals in transformation models is to generalize the
notion of a residual to a wider class of models, namely TMs, and allow for 
all kinds of uninformative censoring and response types.

In Definition~\ref{def:score}, we note that the score residual can be 
interpreted as the score contribution of a newly introduced
intercept parameter, which is constrained to zero. This is equivalent
to formulating a score test for $\alpha = 0$ for a covariate that is
not yet included in the model.
Since $\alpha$ is constrained to zero in the whole procedure, we do not
need to evaluate the model under the alternative hypothesis.
Note also, that we restrict $\alpha$ to an intercept term on the scale
of the transformation function. \citeauthor{farrington2000residuals} gives 
a compelling argument why one should do so. This connection is drawn next.

As an adjustment of Cox-Snell residuals \citep{cox1968general}, 
\citeauthor{lagakos1981residuals} proposes a centered version of Cox-Snell 
residuals.
\citet{barlow1988residuals} later drew the connection to stochastic calculus
and coined the term ``martingale residual'' \citep[see also][]{therneau1990martingale}.
For interval and right censored, as well as exact responses, Lagakos residuals 
have a direct connection to score statistics \citep{farrington2000residuals}.
The setting is based in survival analysis, but the connection
to transformation models will become apparent.
\citeauthor{lagakos1981residuals} starts from a general proportional hazards model where
\begin{align*}
    \lambda(t \given \rx) = \lambda(t) \exp(\rx^\top\shiftparm)
\end{align*}
denotes the hazard function depending on time $t$ and covariates $\rx$,
which can be decomposed into a product of a baseline hazard function
$\lambda(t)$ and the exponential function of a linear predictor in
the covariates $\shiftparm$. The cumulative hazard function is then defined
as
\begin{align*}
    \Lambda(t \given \rx) = \int_0^t \lambda(u \given \rx) du. 
\end{align*}
From there, the cdf can be recovered as
\begin{align*}
    F_T(t \given \rx) = 1 - \exp(-\Lambda(t)\exp(\rx^\top\shiftparm)).
\end{align*}
By definition $\Lambda(t) > 0 \; \forall t$, hence
\begin{align*}
    F_T(t \given \rx) = 1 - \exp(-\exp(\log\Lambda(t) + \rx^\top\shiftparm)),
\end{align*}
which is a transformation model with Minimum-Extreme-Value inverse-link, $F_Z = \pMEV$, and 
the transformation function, $\h(t) = \log\Lambda(t)$, is the log cumulative baseline hazard.
\citeauthor{farrington2000residuals} now assumes that the baseline hazard function belongs to
a family $\calF$ that is closed under scaling by a factor $\gamma > 0$, \ie
\begin{align*}
    \Lambda(t \rvert x) \in \calF \Rightarrow \gamma\Lambda(t \given \rx) \in \calF.
\end{align*}
The Lagakos residual is now defined as
\begin{align*}
    \hat{r}^L = \frac{\partial}{\partial \alpha} \ell 
        \bigg\rvert_{\hat\shiftparm,\hat{S}_0}
\end{align*}
for $\alpha = \log\gamma$ and $\hat{S}_0$ being the estimated survivor curve
under $\rx = 0$, \ie $\hat{S}_0 = \exp(-\hat{\Lambda}(t \given \rx = 0))$.
This translates to the family of log (cumulative) hazard functions being closed under
addition of $\alpha = \log\gamma \in \mathbb{R}$, which is exactly the case for transformation
models. 
This step corresponds to adding and constraining a new intercept 
term to zero on the scale of the transformation function.
Lagakos residuals behave like the usual score statistic, \ie they have mean
zero asymptotically and are asymptotically uncorrelated.

\section{Score residual forms} \label{app:score-resid-forms}

In Section~\ref{sec:bg}, we have seen the equivalence between score and least square
residuals for the normal linear regression model. In Section~\ref{sec:danchor},
we have shown that martingale residuals are score residuals for exact and right censored
observations in a Cox PH transformation model. In the following, we present more 
examples of commonly appearing score residuals.

\begin{example}[Normal distribution with non-linear transformation function]
\label{ex:boxcox}
When choosing $\pZ = \Phi$ and a non-linear transformation function, \eg
\begin{align} \label{eq:nonlintrafo}
    \h(\ry\given\rx) = \h_0(\ry) - \rx^\top\shiftparm,
\end{align}
the score residual of an exact response is
\begin{align*}
    r = \hat\h_0(\ry) - \rx^\top\hat\shiftparm.
\end{align*}
Thus, we arrive at a least-squares residual for the transformed
response. Here, $r \in (-\infty, \infty)$.
\end{example}

\begin{example}[Logistic distribution with non-linear transformation function]
Choosing $\pZ = \expit$ and the same transformation function as
in \eqref{eq:nonlintrafo}, the score residual of an exact response
is
\begin{align*}
    r = 2 \expit(\hat\h_0(\ry) - \rx^\top\hat\shiftparm) - 1.
\end{align*}
Thus, in case the model predicts $\ry$ well 
(\ie $\hat\h_0(\ry) - \rx^\top\hat\shiftparm$ is close to zero, 
analogously to Example~\ref{ex:boxcox}),
the residual will be close to zero too. Note that here $r \in [-1, 1]$.
For a right censored observation $\ry \in (\ubar\ry, \infty)$, the 
score residual is simply the estimated probability
\begin{align*}
    r = \expit(\hat\h_0(\ubar\ry) - \rx^\top\hat\shiftparm) =
    \Prob(\rY \leq \ubar\ry \given \rx; \hat\h_0, \hat\shiftparm),
\end{align*}
and for a left censored observation $\ry \in (-\infty, \bar\ry]$
\begin{align*}
    r = - (1 - \expit(\hat\h_0(\bar\ry) - \rx^\top\hat\shiftparm)) = 
    -\Prob(\rY > \bar\ry \given \rx; \hat\h_0, \hat\shiftparm).
\end{align*}
Again, a good prediction should put little probability mass on
either of the above two events and thus yield a small score residual.
For the right censored case the residual is restricted to $r \in [0, 1]$, 
whereas for the left censored case to $r \in [-1, 0]$, which confirms
the intuition that one can err only in one direction for such responses.
\end{example}

In the examples above, stabilizing the residuals across values of the 
anchors in distributional anchor regression can be expected 
to improve worst-case prediction under perturbations of the anchors.

\section{Computational Details} \label{app:comp}

Anchor TMs, simulation scenarios and visualizations are written in the
\proglang{R} language for statistical computing \citep[Version 3.6.3,][]{pkg:base}.
To implement distributional anchor regression, note that the transformation model
log-likelihood is concave w.r.t. $\parm$, unless all observations are right censored.
The penalty resulting from the quadratic form $\rvec^\top\prm\rvec$ is convex if
$\rvec$, from Definition~\ref{def:score}, is affine or convex \citep{boyd2004convex}.
In case of exact observations in Lm and c-probit, the resulting penalty is convex
and thus solvable by a constrained convex optimizer.
\citet{kook2020regularized} implement regularized transformation models under
elastic net penalties in \pkg{tramnet}, using the domain-specific language optimizer 
\pkg{CVXR} \citep{fu2020cvxr}. From \pkg{tramnet}, we use the TM likelihood implementation.
For the interval censored or right censored models (o-logit, c-logit) we fit
the models using (stochastic) gradient descent (SGD) from package \pkg{TensorFlow}
\citep{tensorflow2015}. The implementation of the interval censored log-likelihood
for ordinal TMs was taken from \citet{kook2020ordinal} and we used SGD with the
Adam optimizer \citep{Kingma2015adam} with learning rate $10^{-3}$, batch size
$250$ and $200$ epochs. Parameters were initialized with the maximum likelihood
estimate for $\xi = 0$ obtained via \code{tram::Polr()} \citep{pkg:tram}.

\section{Invalid Instruments and Shrinkage of Estimates} \label{app:invalid}

In the linear IV setting an instrument is called weak if it has little impact
on the explanatory variables $\rX$.
Consequently, a two-stage least squares procedure will yield homogeneous
predictions for $\rX$ in the first stage, because
\begin{align*}
    \rX \perp \rA \implies \prm \mX \equiv \operatorname{const.}
\end{align*}
This leads to regressing the response on a matrix with constant columns,
making it impossible to separate $\shiftparm$ from an intercept.
Thus, in case of weak instruments, the resulting estimates $\hat\shiftparm$
will shrink towards $0$ as $\xi \to \infty$ when using a causal regularizer.
This explains the effect seen for the BostonHousing2 data in Section~\ref{sec:app}
and makes the diluted causal parameter impossible to study, because it is equal 
to constant 0.
However, intermediary amounts of causal regularization yield a benefit in terms
of worst-case LOEO CV (\cf Figure~\ref{fig:res:bh2}).

\section{Additional simulation results} \label{app:sim}

In this appendix, we report additional simulation results to more clearly substantiate
the empirical findings in the main text.

\subsection{Scenario {\sc nla}}

\begin{figure}[!ht]
    \centering
    \includegraphics[width=0.7\textwidth]{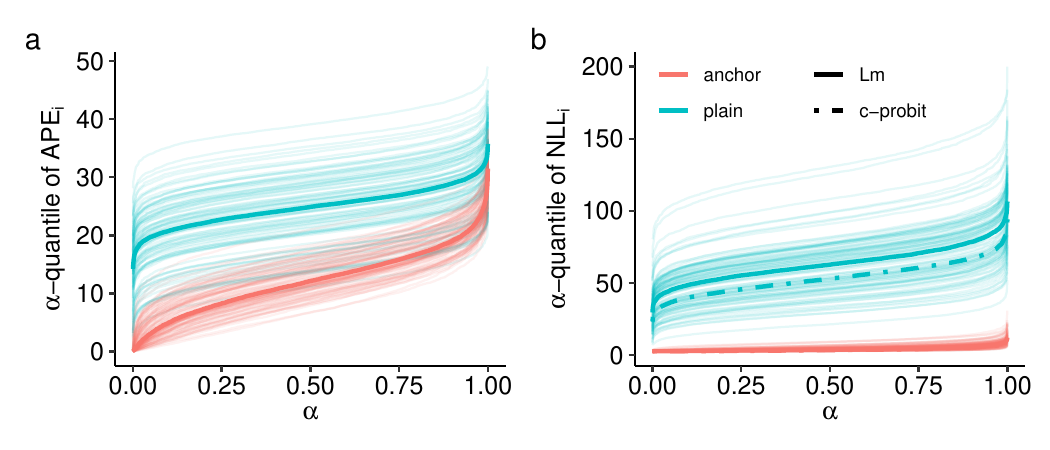}
    \caption{
    Test performance (quantile functions) over 100 simulations for scenario {\sc nla} with 
    the same parameters as in Fig.~\ref{fig:res:nla}, but using a mis-specified
    linear $L_2$ anchor (\textsf{a}) and normal linear anchor transformation (\textsf{b}) model.
    The quantile function of the absolute prediction errors (\textsf{a}) and 
    negative log-likelihood (\textsf{b}) contributions are shown with $\gamma = 13$ and 
    $\xi = 6$, respectively. The point-wise median over all simulations is indicated by a 
    thick line.
    As a reference, we show the point-wise median NLL of the c-probit model from
    Fig.~\ref{fig:res:nla} in \textsf{b}.
    }
    \label{fig:nlalm}
\end{figure}

To show that the DGP for scenario {\sc nla} is sufficiently challenging to warrant
a non-linear anchor regression model, we fit a mis-specified linear $L_2$ anchor regression
with $\gamma = 13$ and normal linear anchor TM with $\xi = 6$ to the data. As mentioned
in the main text, the non-linear conditional expectation function $f(\rx) = \Ex[\rY\given\rX=\rx]$
extends linearly outside the training support and thus linear anchor regression performs
only slightly worse (Fig.~\ref{fig:nlalm}).
In absolute terms, the mis-specified plain Lm performs worse than the c-probit model
(\cf the dot-dashed line in Fig.~\ref{fig:nlalm}\textsf{b}), whereas the anchor regression
models perform similar.

\begin{figure}[!ht]
    \centering
    \includegraphics[width=0.7\textwidth]{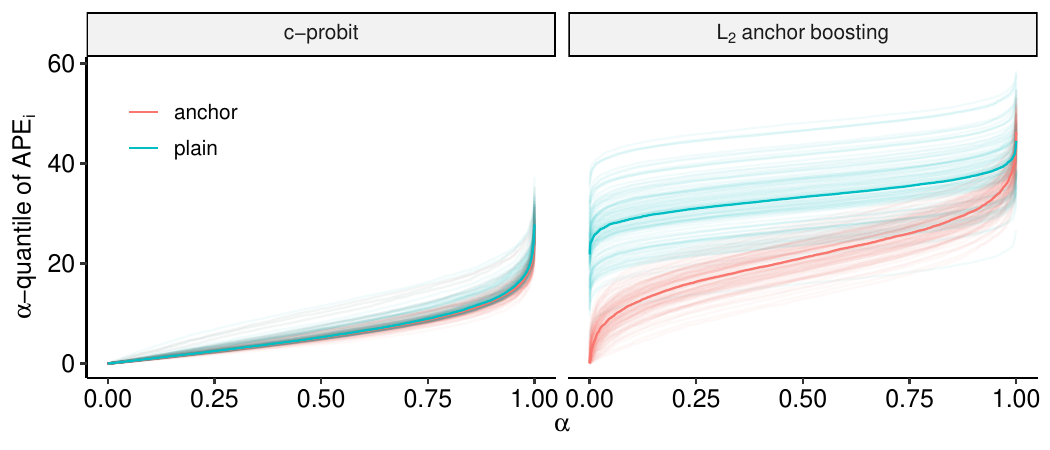}
    \caption{
    Test performance (quantile functions) over 100 simulations for scenario {\sc nla} with 
    the same parameters as in Fig.~\ref{fig:res:nla}, using conditional mean predictions
    from $L_2$ anchor boosting ($\gamma = 13$) and conditional median predictions from a 
    c-probit model ($\xi = 6$).
    The point-wise median over all simulations is indicated by a thick line.
    }
    \label{fig:nlaboosting}
\end{figure}

Point estimates for anchor TMs can be obtained via the conditional quantile function, \eg
the conditional median. However, we strongly recommend using models tailored towards conditional
mean estimation in anchor regression, such as $L_2$ anchor boosting \citep{buehlmann2018invariance}.
In Fig.~\ref{fig:nlaboosting}, the performance of anchor TMs (c-probit, $\xi = 6$) in terms of point 
prediction is compared with anchor boosting ($m_{\text{stop}} = 50$, $\gamma = 13$). Obtaining a
conditional mean prediction in anchor TMs is cumbersome, because it involves numeric integration
over incomplete densities.

The conditional median seems to perform reasonably well, even compared to conditional mean 
prediction in anchor boosting 
(assuming symmetric errors, the two prediction errors can be directly compared).
However, the benefit of using distributional anchor regression is more strongly reflected in
the NLL and not the APE, since the first is a proper scoring rule evaluating a probabilistic
instead of a point prediction.

The performance gain in terms of NLL for c-probit remains the same as in Fig.~\ref{fig:res:nla}.
We refrain from comparing the anchor TM NLL against an anchor boosting NLL for which one could assume
a Gaussian likelihood, because the method is not intended for distributional predictions.

\subsection{Scenario {\sc iv1}}

In a similar vein, to show that the DGP for scenario {\sc iv1} is sufficiently challenging,
we again fit a mis-specified linear $L_2$ anchor regression with $\gamma = 15$ and normal linear 
anchor TM with $\xi = 7$ to the data. 
There is a performance gain in favor of linear anchor regression over the plain linear model for 
both APE and NLL. However, for small quantiles (up until the median), the mis-specified model 
performs worse than the c-probit model (\cf Fig.~\ref{fig:res:iv1}). The mis-specified model
nevertheless exerts robustness properties and shows better performance than the correctly specified
model for quantiles larger than 0.5 (but excluding the maximum).
The DGP in scenario {\sc iv1} can be deemed sufficiently challenging to show the benefits of 
non-linear, distributional extensions of linear $L_2$ anchor regression. The mis-specified models
still improve worst-case prediction, but suffer from worse performance for those observations that
are ``easy'' to predict for the non-linear model.

\begin{figure}[!ht]
    \centering
    \includegraphics[width=0.8\textwidth]{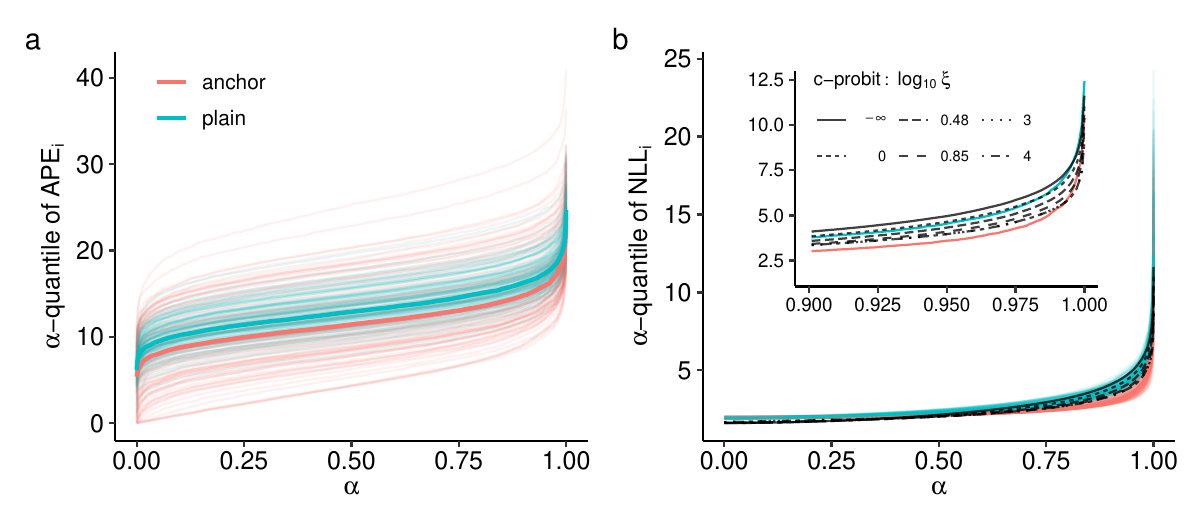}
    \caption{
    Test performance (quantile functions) over 100 simulations for scenario {\sc iv1} with 
    the same parameters as in Fig.~\ref{fig:res:iv1}, but using a mis-specified
    linear $L_2$ anchor (\textsf{a}) and normal linear anchor transformation (\textsf{b}) model.
    The quantile function of the absolute prediction errors (\textsf{a}) and 
    negative log-likelihood (\textsf{b}) contributions are shown with $\gamma = 15$ and 
    $\xi = 7$, respectively. The point-wise median over all simulations is indicated by a 
    thick line. 
    The original point-wise median NLL of the correctly specified c-probit model 
    from Fig.~\ref{fig:res:iv1} is depicted in black.
    }
    \label{fig:iv1lm}
\end{figure}

\subsection{Scenario {\sc iv2}}

\begin{figure*}[!ht]
    \centering
    \includegraphics[width=0.9\textwidth]{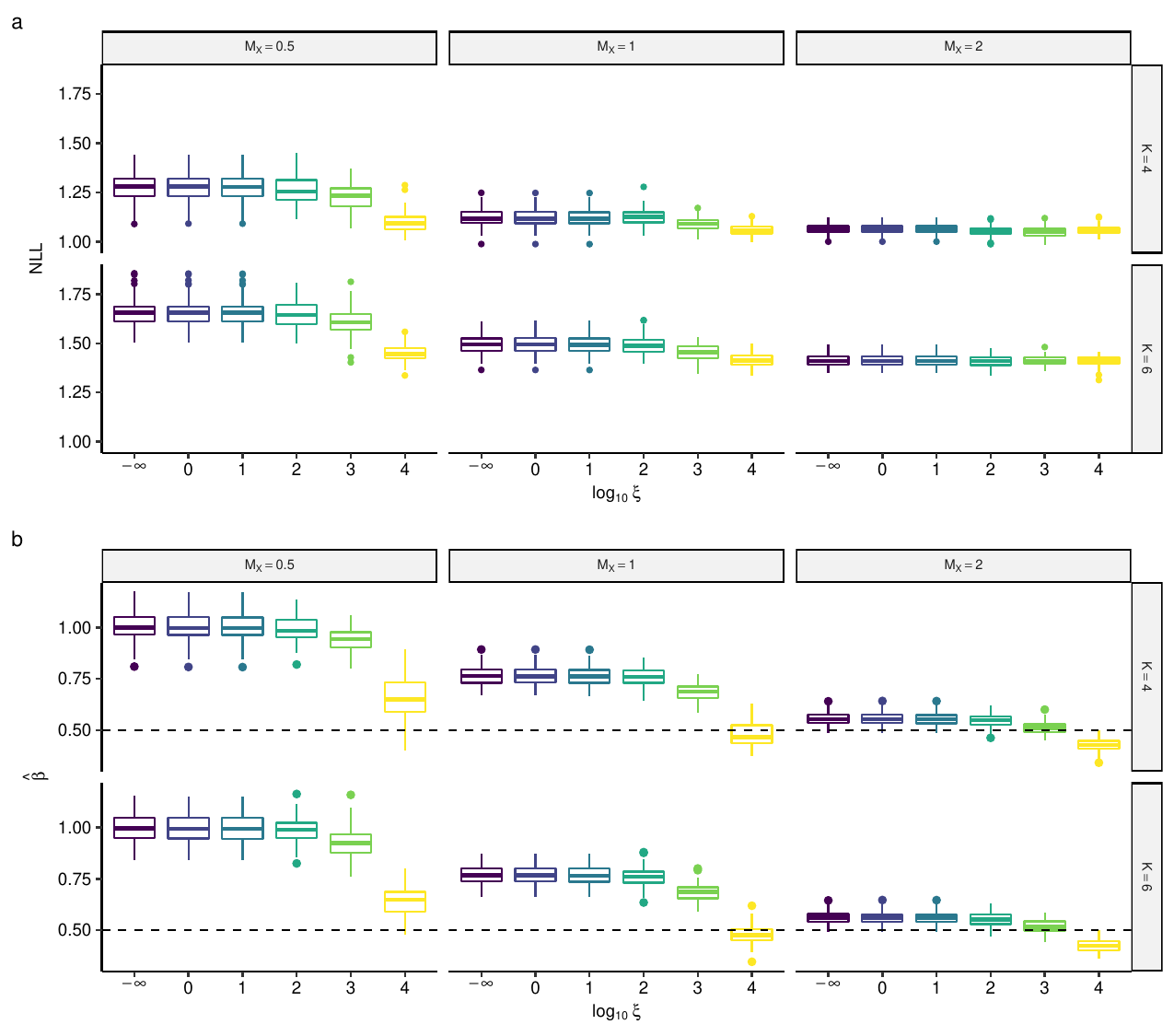}
    \caption{
    Simulation results for scenario {\sc iv2} for $K \in \{4, 6\}$ repeated 100 times.
    \textsf{a}: The average test NLL is displayed for each simulation, varying $M_X$ and
    constant $\pdo(A = 3)$. \textsf{b}: Coefficient estimates for each model, where the
    simulated ground truth ($\beta = 0.5$) is indicated by a dashed line. 
    For details, see Fig.~\ref{fig:res:iv2}.
    }
    \label{fig:iv2k4}
\end{figure*}

Here, we show additional simulation results for scenario {\sc iv2} using a different number
of classes for the ordinal outcome, namely $K = 4$ and $K = 6$. We restrict ourselves to the 
perturbation $\pdo(A = 3)$ (Fig.~\ref{fig:iv2k4}). 
The same main conclusions as for $K = 10$ can be drawn. However, responses
with bounded sample space suffer from a common problem in linear model formulations if
the shift perturbations are overly large. In this case, the response's marginal distribution
becomes extremely skewed towards class $1$ or $K$.

\end{document}